\documentclass[english]{iopart}

\usepackage{graphicx}
\usepackage{url}
\usepackage{iopams} 

\expandafter\let\csname equation*\endcsname\relax
\expandafter\let\csname endequation*\endcsname\relax

\usepackage{amssymb}
\usepackage[latin1]{inputenc}
\usepackage{amsmath}
\usepackage{epsfig}
\usepackage{color}
\definecolor{darkblue}{rgb}{0, 0, .4}

\usepackage[bookmarks]{hyperref}
\hypersetup{
        colorlinks=true,
        linkcolor=darkblue
}

\newcounter{todocounter}

\newcounter{fig}
\begin{document}

\title[Anisotropic correlations]
{\Large \bf The anisotropic Ising correlations as elliptic integrals:
  duality and differential equations }

\vskip 0.3cm

\vspace{.1in}

\author{B. M. McCoy$^\ddag$ and
J-M. Maillard$^\dag$}

\address{$^\ddag$ C N Yang Institute for theoretical Physics, 
State University of New York, Stony Brook, N Y 11794, USA} 

\address{$^\dag$ LPTMC, UMR 7600 CNRS, 
Universit\'e de Paris 6,  
Tour 23, 5\`eme \'etage, case 121, 
 4 Place Jussieu, 75252 Paris Cedex 05, France} 

\vspace{.2in}

E-mail:  mccoy@max2.physics.sunysb.edu, maillard@lptmc.jussieu.fr,

\vskip .01cm

\vskip .3cm 

\vskip .2cm 

{\em Dedicated to A. J. Guttmann, for his 70th birthday.}

\vspace{.1in}

\begin{abstract} We present the reduction of 
the  correlation functions 
of the Ising model on the anisotropic square lattice to 
complete elliptic integrals of the first, second and third kind, 
the extension of Kramers-Wannier duality to anisotropic 
correlation functions, and the linear differential
equations for these anisotropic correlations. More precisely, 
we show that the
anisotropic correlation functions are homogeneous polynomials
of the complete elliptic integrals of the first, second and 
third kind. We give the exact dual transformation matching
the correlation functions and the dual correlation functions.
We show that the linear differential operators annihilating the
general two-point correlation functions are factorised 
in a very simple way, in operators of decreasing orders. 

\end{abstract} 

\vskip .3cm

\noindent {\bf PACS}:
05.50.+q, 05.10.-a, 02.10.De, 02.10.Ox

\vskip .3cm

\noindent {\bf Key-words}:
Dual correlation functions of the anisotropic Ising model, 
high and low temperature correlation functions, 
quadratic relations on the correlation functions,
complete elliptic integrals of the first, second and third kind, 
Kramers-Wannier duality, partial differential equations.

\vskip .2cm 

\section{Introduction}
\label{Introduction}

The two dimensional Ising model
has been the object of penetrating investigations beginning 
with the loop algebraic computation of the free energy in 1944 
by Onsager~\cite{onsager},the
spinor (Fermionic) method of Kaufman~\cite{kaufman} in 1949 , the
correlation computations of Kaufman and Onsager~\cite{ko} also in
1949,  the characterization as
Toeplitz determinants by Montroll, Potts and Ward~\cite{mpw} in 1963, 
the asymptotic behavior for large separations of Wu~\cite{wu1966} in
1966, the Painlev{\'e} III representation in the  scaling
region of  $\, T$ near $\, T_c$ by Wu, McCoy, 
Tracy and Barouch~\cite{wmtb} in 1976   
and the Painlev{\'e} VI representation of the diagonal 
correlation functions by Jimbo and Miwa~\cite{jm1} in 1981. We know 
more about the correlation functions of the Ising model than
 any other system and these correlation functions have
inspired developments ranging from conformal field 
theory to random matrices.

 Nevertheless there are still many questions
which remain unsolved. In this note we investigate the
reduction of the anisotropic correlation functions to 
{\em homogeneous polynomials}
in the three kinds of complete elliptic integrals, the 
implications of the Kramers-Wannier duality~\cite{kw} 
and the linear differential equations satisfied by 
the anisotropic correlation functions. 

In section \ref{nearest} we review
the reduction of the nearest neighbor correlation to the 
form obtained by Onsager~\cite{onsager}. This leads to an
identity on complete elliptic integrals 
of the third kind.
In section \ref{generalcorr} we extend this complete elliptic 
integral representation to all correlations.
In section \ref{duality} we find a representation 
of the Kramers-Wannier duality on the complete elliptic 
integral of the first, second and third kind,
which actually transforms the correlation functions into
the dual correlation functions.
We conclude in section \ref{lindiff} with a discussion of 
the linear differential equations for the correlations.

\vskip .1cm

\section{The nearest neighbor correlation}
\label{nearest}

The two dimensional Ising model on a square lattice is 
defined by the interaction energy
\begin{eqnarray}
\label{defcalE}
\hspace{-0.95in}&&  \, \,  \quad  \quad  
\quad   \quad  \quad    \quad   \quad    \quad  
{\mathcal E}\,\,  = \,\,  \, 
-\sum_{i,j} \, \, (E_v \cdot \, \sigma_{i,j}\,\sigma_{i+1,j}
\, \,  +E_h \cdot \,\sigma_{i,j}\, \sigma_{i,j+1}), 
\label{interaction}
\end{eqnarray}
where $\, \sigma_{i,j}\, =\, \pm 1$ is the spin at row 
$\, i$ and column $\, j$ and the sum is over all values 
$\pm \, 1$ for all spins in a  
lattice of $\, L_v$ rows and
$\, L_h$ columns with either cylindrical or toroidal boundary
  conditions. The partition function on the 
$\, L_v\, \times\,  L_h$ lattice at
  temperature $\, T$ and 
the free energy in the thermodynamic limit, are defined as
\begin{eqnarray}
\label{defZ}
\hspace{-0.95in}&&  \, 
Z(\beta;L_v,L_h)\, = \,\, \, \, 
\sum_{\sigma=\pm 1} e^{-\beta {\mathcal E}}, \quad \quad 
F\,\,  =\, \,\,  
-k_B \; T \, \lim_{L_v,L_h\rightarrow \infty}\frac{1}{L_v L_h} \, \ln Z(\beta;L_v,L_h). 
\end{eqnarray}

The result of this computation is the famous double integral formula
for the free energy
\begin{eqnarray}
\hspace{-0.95in}&&  \, \,  \quad  \quad  \quad    \quad  
-\beta F \, \, \, = \, \, \, \, \,  \ln 2 \,\, \,\,  
+\frac{1}{2 \, (2 \pi)^2} \, \int_0^{2\pi}  \,
d\theta_1 \cdot \, \int_0^{2\pi}d\theta_2 \cdot \, 
  \ln\{\cosh 2\beta E_v \, \cosh 2\beta E_h
\nonumber\\
\label{free}
\hspace{-0.95in}&& \quad \quad \quad \quad  \quad
 \quad  \quad \quad \quad  \quad  
\, \, -\sinh 2\beta E_v \, \cos \theta_1 
\, -\sinh 2\beta E_h \, \cos \theta_2\}, 
\end{eqnarray}
where $\,  \beta= \,  1/k_BT$ ($k_B$ being Boltzmann's constant).
This free energy has a singularity of the form
$ \, (\beta-\beta_c)^2 \cdot \ln(\beta-\beta_c)^2$,  
at the critical value $ \, \beta_c = \, 1/k_BT_c$,  defined by:
\begin{eqnarray}
\hspace{-0.95in}&&  \, \,  \quad   \quad  
 \quad   \quad   \quad   \quad   \quad  \quad   
\sinh2\beta_cE_v \cdot \, \sinh 2\beta_cE_h\,\, = \,\,\, \pm \, 1. 
\end{eqnarray}

We begin our discussion of Ising correlation functions with
the nearest neighbor row and column correlation functions which 
can be obtained from the free energy as
\begin{eqnarray}
\label{intenergy}
\hspace{-0.95in}&&  \, \,  \quad  
\quad  \quad   \quad  \quad  \quad  
u\,\,  =\, \, \, \frac{\partial\beta F}{\partial \beta} \, 
=\,\,\,  -E_h \cdot  \, \langle\sigma_{0,0}\sigma_{0,1}\rangle
\,\,  \, -E_v \cdot  \, \langle\sigma_{0,0}\sigma_{1,0}\rangle. 
\end{eqnarray}
Thus by differentiating (\ref{free}) and doing one of 
the integrals by closing a contour on  poles we find 
\begin{eqnarray}
\hspace{-0.95in}&&  \, \,  \quad      \quad 
C(0,1) \, \, = \, \,  \, \,
 \langle\sigma_{0,0}\sigma_{0,1}\rangle  
\,\, = \,  \, \, \,
\frac{1}{2\pi}\int_{0}^{2\pi}  \, 
d\theta \cdot \, 
\left[\frac{(1-\alpha_1 e^{i\theta})(1-\alpha_2 e^{-i\theta})}
{(1-\alpha_1e^{-i\theta})(1 -\alpha_2 e^{i\theta})}\right]^{1/2}, 
\label{c01}
\end{eqnarray}
with
\begin{eqnarray}
\hspace{-0.95in}&&  \, \,  \quad  \quad  \quad  \quad  
\alpha_1\,\,  =\,\, \,  
z_h \cdot  \,\frac{1-z_v}{1+z_v}
\,\, \,   = \, \, \, \tanh \beta E_h \cdot \, e^{-2\beta E_v}, 
\label{a1} 
\\
\hspace{-0.95in}&&  \, \,  \quad  \quad  \quad  \quad  
\alpha_2\,\,  =\,\,\,   
 {{1} \over {z_h}} \cdot \, \frac{1-z_v}{1+z_v}
\, \,\,   =\, \, \, \coth \beta E_h \cdot \, e^{-2\beta E_v}, 
\label{a2}
\end{eqnarray}
where $  \, \,z_v= \, \, \tanh \beta E_v \,\, $ and 
$ \,\, z_h= \, \tanh \beta E_h$.

The result (\ref{c01}) can be reduced to complete elliptic 
integrals of the first, second and third kind defined as
\begin{eqnarray}
\label{elliptick}
\hspace{-0.95in}&&  \, \,  \quad\quad    \quad  \quad  \quad  
{\tilde K}(k) \, \,  = \,  \,  \, 
\frac{2}{\pi} \cdot \,
 \int_0^{\pi/2}\frac{d\phi}{(1-k^2\sin^2\phi)^{1/2}},
\\
\label{elliptice}
\hspace{-0.95in}&&  \, \,  \quad  \quad  \quad  \quad  \quad  
{\tilde E}(k) \, \,  = \, \,   \, 
\frac{2}{\pi} \cdot \, 
\int_0^{\pi/2} \, d\phi \cdot \, (1-k^2\sin^2\phi)^{1/2},
\\
\label{elliptic3}
\hspace{-0.95in}&&  \, \,  \quad \quad   \quad  \quad  \quad  
{\tilde \Pi}(n,k) \, \,  = \, \, \, 
\frac{2}{\pi} \cdot \, \int_0^{\pi/2} \, 
 \, \frac{d\phi}{(1-n \sin^2\phi)(1-k^2\sin^2\phi)^{1/2}}, 
\end{eqnarray}
where we have chosen a normalization such that at 
$\, k\, = \, n= \, 0$ the three elliptic integrals are 
unity instead of $ \, \pi/2$. We also introduce the 
{\em modulus of the elliptic functions} 
parametrizing the model:
\begin{eqnarray}
\label{defkk}
\hspace{-0.95in}&&  \, \,  \quad  \quad 
 \quad   \quad  \quad   \quad  
k \, \,= \, \, \, \sinh 2 \beta E_v \cdot \, \sinh 2 \beta E_h
\, \, =\, \, \,
 \frac{\alpha_2^{-1}-\alpha_1}{1 \, -\alpha_1\alpha_2^{-1}},
\\
\label{defkkminus}
\hspace{-0.95in}&&  \, \,  \,  \,  \quad  
\quad   \quad  \quad   \quad  
  k_< \, \,= \, \, \,  {{1} \over {k}} 
\, \,= \, \, \, 
(\sinh 2 \beta E_v \, \sinh 2\beta  E_h)^{-1} 
\, \,=\, \, \,
\frac{\alpha_2-\alpha_1}{1-\alpha_1\alpha_2}.
\end{eqnarray}

This reduction is carried out in~\cite{book} by two different 
methods which lead to two, at first sight, {\em different} 
expressions. Perhaps the most straightforward reduction for 
$ \, T \, < \, T_c~(\alpha_2 \, < \, 1) $
is to set $ \, e^{i\theta} \, = \, \zeta$  in (\ref{c01}), 
and, then, use the substitution
\begin{eqnarray}
\label{variablechange}
\zeta \,\, \,= \, \,\, \,
  \frac{\alpha_1 \cdot \,(\alpha_1^{-1}-\alpha_2)
 \,  \, +\alpha_1^{-1}\cdot \, (\alpha_2 -\alpha_1) \cdot \, \sin^2\phi}
{\alpha_1^{-1}-\alpha_2 \,  \, +(\alpha_2 -\alpha_1) \cdot \, \sin^2\phi}, 
\end{eqnarray}
to obtain:
\begin{eqnarray}
\label{c01mbad}
\hspace{-0.95in}&&  \, \,  \,   \,    \quad   \quad  
C_<(0,1)\, \,\, = \, \, \, \, \, 
\frac{\alpha_1 -\alpha_1^{-1}}{\alpha_2^{-1} -\alpha_1} \cdot \, 
\{{\tilde K}(k_<) \,  \, 
-(1+\alpha_1\alpha_2^{-1}) \cdot \, {\tilde\Pi}(-\alpha_1k_<,k_<)\}, 
\end{eqnarray} 
A similar computation in~\cite{book} for 
$ \, T \, > \, T_c~(\alpha_2 \, > \, 1)$ 
gives:
\begin{eqnarray}
\hspace{-0.95in}&&  \, \,  \, \,\, \, \, \,
C_>(0,1) \, \,  \, = \, \, \, \, 
\frac{\alpha_1-\alpha_1^{-1} }
{1-\alpha_1\alpha_2^{-1}} \cdot \, 
\{{\tilde K}(k)\,\,  
-(1+\alpha_1\alpha_2^{-1}) \cdot \,{\tilde \Pi}(-\alpha_1k, \, k)\}. 
\label{c01pbad}
\end{eqnarray}

This form, however, is not particularly transparent and 
a more elaborate reduction, given in \cite{book}, gives\footnote[1]{Removing 
a misprinted factor of $\, \alpha_2$ in (3.70) on page 97 of~\cite{book}.} 
a form first obtained by Onsager~\cite{onsager}. Using 
the notation
\begin{eqnarray}
\hspace{-0.95in}&&  \, \,  \quad \quad  
\quad   \quad  \quad  \quad  \quad   
s_h \,\,  = \, \, \, \sinh 2\beta E_h, 
\quad \quad \quad \quad 
s_v \, \, =\, \, \, \sinh 2\beta E_v, 
\end{eqnarray}
the result for $ \, T \, < \, T_c$ is
\begin{eqnarray}
\label{Clow01}
\hspace{-0.95in}&&  \, \,  \quad  
C_<(0,1) \,\,  = \,\,  \,\, 
  {\sqrt{1+s^2_v}} \cdot\, s_v 
\cdot \,  s_h^{-2} \cdot \, \{(1+s_h^2) \cdot \, 
{\tilde \Pi}(-s_v^{-2}, \, s_v^{-1}s_h^{-1}) \, \,  \, 
-{\tilde K}(s_v^{-1} \, s_h^{-1})\}
\nonumber\\
\label{onscorm}
\hspace{-0.95in}&&  \, \,  \quad  \quad  \quad 
 = \, \, \, 
 {\sqrt{1+\nu k_<}} \cdot \,
 \{(1+k_</\nu) \cdot \, {\tilde \Pi}(-\nu k_<, \, k_<)
 \, \,  \, -(k_</\nu) \cdot \, {\tilde  K}(k_<)\}, 
\end{eqnarray}
 and for $\,T\, > \, T_c$
\begin{eqnarray}
\label{onscorp}
\hspace{-0.95in}&&  \, \,  \quad   \quad 
C_>(0,1)\, \, = \, \, \, \, 
{\sqrt{1+s^2_v}}\cdot  \, s_h^{-1} \cdot  \,
\{(1+s_h^2) \cdot \, {\tilde\Pi}(-s_h^2, \, s_vs_h)
 \,  \, \, -{\tilde K}(s_vs_h)\},
\nonumber\\
\hspace{-0.95in}&&  \, \,  \quad  \quad  
\quad  \quad  \quad  \quad 
= \,\,\,
\frac{1}{\nu} 
\cdot  \, {\sqrt{1+\nu/k}} \cdot  \, \{(1+\nu k) \cdot \,
{\tilde \Pi}(-\nu \,  k, \, k) \,  \, \, -{\tilde K}(k)\}, 
\end{eqnarray}
where the anisotropy $ \, \nu$ is defined as:
\begin{eqnarray}
\label{nudef}
\hspace{-0.95in}&&  \, \,  \,  \quad \quad  
\quad   \quad  \quad   \quad 
\nu \,\,\, =\,\,\, \frac{s_h}{s_v}
\, \,\,  =\,\, \, 
\frac{4\,  \alpha_1\,   \alpha_2}{(\alpha_2\,  -\alpha_1) 
\cdot \, (1\,  -\alpha_1\alpha_2)}. 
\end{eqnarray}
One also has: 
\begin{eqnarray}
\hspace{-0.95in}&&  \,  \,  \,  \quad \quad 
\nu \cdot \, k_<\,\,  =\,\, \, 
 \frac{4\alpha_1\alpha_2}{(1-\alpha_1\alpha_2)^2}
\,\,  =\, \, \,  s_v^{-2},
\quad \quad \, \, \,   
\nu \cdot \,   k\,\,    =\, \, \,  
  \frac{4\alpha_1\alpha_2}{(\alpha_2-\alpha_1)^2}
\, \,  = \, \, \, s_h^2.
\end{eqnarray}

We note that the high temperature correlation (\ref{onscorp}) 
is obtained from the low temperature correlation (\ref{onscorm}) 
by the substitution
\begin{eqnarray}
\label{dual1}
\hspace{-0.95in}&&  \,  \,  \,  \quad \quad  \quad \quad 
{\tilde K}(s_v^{-1} s_h^{-1})
\quad \quad  \longrightarrow \quad \quad 
s_v s_h \cdot \, {\tilde K}(s_v s_h),
\\
\label{dual2}
\hspace{-0.95in}&&  \,  \,  \,  \quad \quad  \quad \quad 
{\tilde \Pi}(-s_v^{-2}, \, s_v^{-1}s_h^{-1})
\quad \quad   \longrightarrow \quad \quad 
s_v s_h \cdot \, {\tilde \Pi}(-s_h^2, \, s_vs_h).
\end{eqnarray}

The expressions (\ref{c01mbad}) and (\ref{c01pbad}) look 
quite different from (\ref{onscorm}) and 
(\ref{onscorp}). Nevertheless {\em they are actually equal}.
Equating the two forms of the low temperature correlation we
obtain an identity on elliptic integrals 
of the third kind:
\begin{eqnarray}
\label{newident1}
\hspace{-0.95in}&&  \, \quad \quad    
(z_h^2 z_v^2 \, +z_h^2 \, -z_v^2 \, -4z_v \, -1) 
\cdot \,  {\tilde K}(k_<)
\nonumber\\
\hspace{-0.95in}&&  \, \quad \quad   
-2 \cdot \, (z_vz_h+z_v-z_h+1) 
\cdot \, (z_vz_h-z_v-z_h-1) \cdot \, 
{\tilde \Pi}\Bigl(-\frac{(1-z_h^2)(1-z_v)^2}{4z_v}, \, k_<\Bigr)
\nonumber\\
\hspace{-0.95in}&&  \, \quad \quad 
- \, (1+z_v^2) \cdot \, (1+z_h^2) \cdot \, 
{\tilde\Pi}\Bigl(-\frac{(1-z_v^2)^2}{z_v^2}, \, k_< \Bigr) 
\, \, \, = \, \,  \,  \, \, 0, 
\end{eqnarray}
where we have used 
\begin{eqnarray}
\hspace{-0.95in}&&  \,  \quad \quad\quad 
\quad \quad \quad \quad
k_<\, \, = \, \, \, 
\frac{(1-z_v^2)\cdot \, (1-z_h^2)}{4z_vz_h},
\\
\hspace{-0.95in}&&  \, \quad \quad \quad \quad \quad
\frac{(1-z_v^2)^2}{4z_v^2}
\, \,= \, \, \, s_v^{-2}, \, \quad \quad \quad
\frac{(1-z_h^2)(1-z_v)^2}{4 \, z_v}
\, \, =\, \, \, \alpha_1 \cdot \, k_< .
\end{eqnarray}
If we set 
$\, z\,\, = \, \,\, -\alpha_1 \cdot \, k_< \, $, 
we may verify that
\begin{eqnarray}
\hspace{-0.95in}&&  \, \quad \quad 
\quad \quad \quad \quad \quad
-s_v^{-2}\, \, = \, \, \, \, 
4 \cdot \, k_<^2 \cdot \, 
\frac{ z  \cdot \, (z-1) 
\cdot \, (z \, -k^2)}{(z^2-k^2)^2},
\end{eqnarray} 
and, thus, the identity (\ref{newident1}) may be 
rewritten as:
\begin{eqnarray}
\label{piidentity}
\hspace{-0.95in}&&  \, \quad  \quad 
4 \cdot \,  (z-1) \cdot \, (z^2-k_<^2) 
\cdot \, (z-k_<^2) \cdot \, {\tilde\Pi}(z,k_<)
\nonumber\\
\hspace{-0.95in}&&  \, \quad  \quad 
+(z^2+k_<^2-2z) \cdot \, (z^2+k_<^2-2k_<^2z) \cdot \,
{\tilde \Pi}\Bigl(4  \cdot \, k_<^2  \cdot \, \frac{ z  \cdot \,
 (z-1)  \cdot \,(z-k_<^2)}{(z^2 \, -k_<^2)^2}, \, k_<\Bigr)
\nonumber\\
\hspace{-0.95in}&&  \, \quad  \quad \quad \quad 
\, \,- \,  
(z^2-k_<^2) \cdot \, (z^2 -2z -2k_<^2z +3k_<^2) \cdot \, {\tilde K}(k_<)
 \,  \, \,  \,  = \, \, \,  \,  \, 0.
\end{eqnarray}
Performing series\footnote[1]{ For instance series expansions in $\, t$
of the lhs of (\ref{piidentity}) with 
$\, z \, = \, \, (2\, +3\, i)\cdot \, t^3$, $\, k_< \, = \, \, t^2$,
and, conversely, see that {\em another identity} takes place 
when $\, z$ and $\, k_<$ are small but $\, |z/k_<| >>  1$.} 
expansions of the lhs of (\ref{piidentity})
one can check that this identity is valid\footnote[2]{However, 
this identity addressing {\em two complex variables}, it
is difficult to find what is precisely the domain of validity
of this identity in the {\em two complex variables} $\, z$ and $\, k_<$, 
the conditions $\,z \, = \, \pm \, k_<$ certainly playing some role, as 
can be seen on the denominator of transformation (\ref{previous}).} 
when $\, z$ and $\, k_<$ 
are small ({\em not necessarily real}) and such 
that $\, |z/k_<| <<  1$. A  proof of 
identity (\ref{piidentity}) is given in \ref{proof}.

We note that the transformation
\begin{eqnarray}
\label{previous}
\hspace{-0.95in}&&  \, \quad \quad \quad  
\quad  \quad \quad \quad
z \quad \longrightarrow \quad \quad  
4 \cdot \, k_<^2 \cdot \,\frac{z \cdot \, (z-1) 
\cdot \, (z-k_<^2)}{(z^2 \, -k_<^2)^2}, 
\end{eqnarray}
occurring in (\ref{piidentity}), 
is of {\em infinite order}. If one writes the previous 
{\em infinite order}\footnote[5]{Iterating $\, N$ times 
the rational transformation (\ref{previous}), one gets 
rational transformations of the form 
$\, z \rightarrow \, $
$R_N(z,k) \, = \, \, 4^N \cdot \,  z \cdot \, P_N(z, \, k)/Q_N(z, \, k)$,
where $\,P_N(z, \, k)$ and $\, Q_N(z, \, k)$ are 
polynomials of degree in $\, z$ growing, for generic values of $\, k$, 
like $\,4^N$
(for $\, k= \, 1$ these degrees grow like $\, 2^N$). } (rational) 
transformation (\ref{previous}) in terms 
of $\, k \, = \, \, 1/k_<$, it reads:
\begin{eqnarray}
\label{next}
\hspace{-0.95in}&&  \, \quad \quad \quad  
\quad  \quad \quad \quad
z \quad \longrightarrow \quad \quad  
4 \cdot \,\frac{z \cdot \, (1-z) 
\cdot \, (1 \, -k^2 \, z)}{(1 \, -k^2 \, z^2)^2}, 
\end{eqnarray}
where one recognizes, immediately, the doubling transformation,
$\, \theta \, \rightarrow \, \, 2 \, \theta$,  on the 
square of the elliptic sinus, $\, z \, = \, \, sn(\theta, \, k)^2$: 
\begin{eqnarray}
\label{doubling}
\hspace{-0.95in}&&     \quad \quad        \quad    
 sn(\theta, \, k)^2 \, \, 
\quad  \longrightarrow  \, \, \,  \quad   \quad  
sn(2 \, \theta, \, k)^2  \, \,= \, \, \, \,
\nonumber \\
\hspace{-0.95in}&&    \quad   \quad    \quad       \qquad  
\, \, \, \, = \, \, \, \,
 4 \cdot  \, {{  sn(\theta, \, k)^2 \cdot \, 
(1\, - sn(\theta, \, k)^2) \cdot \, 
(1\, - \, k^2 \cdot \, sn(\theta, \, k)^2) 
} \over {(1  \, -k^2 \cdot \, sn(\theta, \, k)^4 )^2 }}.
\end{eqnarray}
The interpretation of this identity as a 
doubling transformation suggests to recall addition formulae
on the Jacobi's Zeta function~\cite{Hancock,Whittaker} (a logarithmic 
derivative of the Jacobi theta function), which is closely related 
to the ratio of the complete elliptic integral of the third kind 
by  the complete elliptic integral of the first kind, like, 
for instance, the relation\footnote[9]{See, for instance, equations 
(64), (65) and (67) in~\cite{Chan}, where the parameter
$\, a$ is {\em not required} to be a rational multiple 
of a period~\cite{Chan} (the rational cases: 
see, for instance, (69) in~\cite{Chan}). We thank J.H.H. Perk for 
reminding us, after completion of this work, the text following (66)
in~\cite{Chan} discussing how the $\Pi_1$ can be 
reduced in rational cases.  } 
\begin{eqnarray}
\label{Zeta}
\hspace{-0.95in}&& 
Z(u+a, \, k) \, \, = \, \, \, \, 
Z(u, \, k)\,\,\,   + \, Z(a, \, k)\,  \, \, 
-k^2 \cdot \, sn(u, \, k) \cdot \, sn(a, \, k) \cdot \,  sn(u+a, \, k).
\end{eqnarray}
After some straightforward calculations one can interpret identity
(\ref{piidentity}) as the $\, u \, = \, a$ limit of identity 
(\ref{Zeta}).  

\vskip .2cm 

\subsection{Isotropic limit}
\label{isolim}

 To obtain the isotropic limit, where $\, \nu= \,  \, 1$, 
we note that elliptic integrals of the third 
kind obey the identity~\cite{cayley}:
\begin{eqnarray}
\label{thirdident}
\hspace{-0.95in}&&  \, \quad \quad  \quad  \quad 
{\tilde \Pi}(-\nu \, k,\, k) \, \, \, 
+{\tilde \Pi}(-\frac{k}{\nu},\, k) \, \, \, 
=\, \, \, \, \,  {\tilde K}(k)\,  \,\, \,
+\left[(1+\nu\, k)(1+\frac{k}\nu)\right]^{-1/2}. 
\end{eqnarray}
Thus, when $ \, \nu  \, = \, 1$, we have
\begin{eqnarray}
\label{identity}
\hspace{-0.95in}&&  \, \quad \quad  
\quad  \quad  \quad  \quad  \quad 
{\tilde \Pi}(-k, \,k)\,\, \,  = \, \, \,\, \, 
 \frac{1}{2} \cdot \, {\tilde K}(k) \,\, \,  
+\frac{1}{2} \cdot \, {{1}\over {1+k}}, 
\end{eqnarray}
which can be used, in (\ref{onscorm}) and (\ref{onscorp}),
to eliminate the elliptic integral of the third kind to find 
for $\, T\, <\, T_c$, in the isotropic limit where 
$\, s_v\, =\, s_h\, =\, s$, that
\begin{eqnarray}
\label{isoc01m}
\hspace{-0.95in}&&  \, \,  \quad  \quad   \quad  \quad \quad 
C_<(0,1)\,\,  =\,\, \,  \, 
(1+k_<)^{1/2} \cdot \, \frac{1}{2}\{(1-k_<) \cdot \, {\tilde K}(k_<)+1\}
\nonumber\\
\hspace{-0.95in}&&  \, \quad \quad \quad 
\quad  \quad   \quad \quad  \quad 
= \,\, \,  (1+s^{-2}) \cdot \, \frac{1}{2}\{(1-s^{-2})
 \cdot \, {\tilde K}(s^{-2})+1\}, 
\end{eqnarray} 
and for $\, T\, > \, T_c$:
\begin{eqnarray}
\hspace{-0.95in}&&  \, \quad \quad  \quad \quad \quad 
C_>(0,1) \, \, = \, \, \,\,
 (1+1/k)^{1/2} \cdot  \,\frac{1}{2}\{(k-1) \cdot \, {\tilde K}(k)+1\}
\nonumber\\
\hspace{-0.95in}&&  \, \quad \quad \quad  \quad 
\quad \quad \quad \quad   \quad 
= \, \, \, 
(1+s^{-2})^{1/2} \cdot \, \frac{1}{2}\{(s^2-1) 
\cdot \, {\tilde K}(s^2)+1\}.
\end{eqnarray}
For $ \, T \, > \, T_c$ as $ \, \,k \, \rightarrow \,  0 \, $ 
we have 
$\,\,\, C_>(0,1)\, \,= \, \, \, k^{1/2}/2\, \, +O(k^{3/2})$.

\vskip .1cm

\section{The general correlation function 
  $ \, C(M,N) =\,  \langle \sigma_{0,0}\; \sigma_{M,N}\rangle$}
\label{generalcorr}

There are two methods which can be used to compute the Ising
correlation functions\footnote[1]{The correlation 
functions of the Ising model are defined as usual, 
see, for instance, equation (14) in~\cite{mpw}.} 
$ \, C(M,N)$: either in terms of their 
representation as determinants~\cite{mpw}, or from their 
{\em quadratic recursion relations}~\cite{mwdiff1,perk,mwdiff2,mpw1981,ayp}.

\subsection{Determinantal representation}
\label{determin}

The next simplest correlations to study, after the nearest
neighbor, are the row $ \, \langle \sigma_{0,0}\;\sigma_{0,N}\rangle$ 
and the diagonal
$ \, \langle \sigma_{0,0}\;\sigma_{N,N}\rangle$ correlation functions. 

The row correlation function 
$\langle \sigma_{0,0}\; \sigma_{0,N}\rangle$ 
can be expressed as an $ \, N \, \times \, N$ Toeplitz 
determinant~\cite{mpw,book}  
\begin{eqnarray}
C(0,N) \,  \, = \, \,  \,  \, 
\begin{array}{|llll|}
{a}_0&{a}_{-1}&\cdots&{a}_{-N+1}\\
{a}_1&{ a}_0&\cdots&{a}_{-N+2}\\
\vdots&\vdots&&\vdots\\
{a}_{N-1}&{a}_{N-2}&\cdots&{a}_0
\end{array}
\label{detdn}
\end{eqnarray}
with
\begin{eqnarray} 
\label{dn2}
\hspace{-0.95in}&&  \, \quad \quad \quad 
\quad  \quad \quad  \quad \quad
a_n \, \,  = \, \,  \, 
{1\over 2\pi} \cdot  \, \int_{0}^{2\pi}  \, d\theta \cdot  \,
e^{-in\theta} \cdot \, w(e^{i\theta}), 
\end{eqnarray}
where the generating function $ \, w(e^{i\theta})$ is
\begin{eqnarray}
\label{dn3}
\hspace{-0.95in}&&  \, \quad \quad \quad 
\quad \quad \quad \quad \quad
w(e^{i\theta})\, \, = \, \, \,
 \left[ {(1-\alpha_1e^{i\theta})(1-\alpha_2e^{-i\theta})\over
(1-\alpha_1e^{-i\theta})(1-\alpha_2e^{i\theta})}\right]^{1/2}, 
\end{eqnarray}
 with $ \, \alpha_1$ and $ \, \alpha_2$ given by 
(\ref{a1}) and (\ref{a2}).
We note, when both $ \, \alpha_1$ and $ \, \alpha_2$ 
are real, that $ \, a_n$ is also real.

For the diagonal correlation $ \, C(N,N)$ we may consider a 
triangular lattice by adding to the square lattice bonds 
of strength $ \, E_d$ connecting sites $(M, \, N)$ with 
sites $ \, (M+1, \, N+1)$, obtaining a determinental 
representation by using a straightline path from 
$\, (0, \, 0)$ to $\, (N, \, N)$ and then setting 
$\,E_d\,= \, 0$. This results in $ \, C(N, \, N)$ being 
given by the determinant (\ref{detdn}) with 
\begin{eqnarray}
\label{dn4}
\hspace{-0.95in}&&  \, \quad \quad \quad \quad \quad \quad \quad
\alpha_1 \, = \,\, 0, \quad \quad \quad \quad 
\alpha_2 \, = \, \, \,  \, \, \, 
(\sinh 2\beta E_v \, \sinh 2\beta E_h)^{-1}. 
\end{eqnarray}

The matrix elements of the diagonal  
correlation function $ \, C(N,N)$ 
are directly seen to be hypergeometric functions which, 
in turn, are expressed as linear combinations of the elliptic 
integrals $ \, {\tilde K}$ and $ \, {\tilde E}$.
For $ \, T \, < \, T_c$ 
\begin{eqnarray}
\hspace{-0.95in}&&  \, \quad \quad \quad \quad \quad \quad
a_0\, \,  = \, \, \,  {\tilde E(k_<)}, 
\end{eqnarray}
and for $ \, T \, > \, T_c$:
\begin{eqnarray}
\hspace{-0.95in}&&  \, \quad \quad \quad \quad \quad \quad
a_0 \,\,  = \, \, \,  \, 
   {{ {\tilde E}(k)} \over { k}} 
\,\,\, - {{1-k^2} \over {k}} \cdot \,  {\tilde K}(k).
\end{eqnarray}

The matrix elements of the {\em row} correlation function
are all expressed 
as linear combinations of the complete elliptic integrals 
${\tilde K}(k),~{\tilde  E}(k),~{\tilde \Pi(s_v^{-2},k)}$ 
and $ \, {\tilde \Pi(z_h^2,k)}$  
by use of the change of variables (\ref{variablechange}) 
and the identity (\ref{newident1}).
For example, 
for $ \, T \, > \, T_c$ with $ \, k \, = \, s_v \, s_h$,  
we have
\begin{eqnarray}
\hspace{-0.97in}&&   \quad \quad \quad \quad
a_{>0} \, \,= \, \, \,
 {\sqrt{1+\nu/k}} \cdot \, 
\{(1+\nu k) \cdot \, 
{\tilde \Pi}(-\nu k,k) \, \, -{\tilde K}(k)\},
\nonumber\\
\hspace{-0.97in}&&  \quad \quad \quad \quad
a_{>\mp1} \, \,= \, \, \,  \, 
\frac{1}{(1+k\nu)^{1/2} \,  \pm \, 1} \cdot \, \{{\tilde E}(k)
\nonumber\\
\hspace{-0.97in}&&  \, \quad \, \, \quad \quad \quad \quad \quad
 \pm \, (1+k\nu)^{1/2} \cdot  \, 
[(1+k/\nu) \cdot  \, 
{\tilde \Pi}(-k\nu,k) \, \, - k/\nu \cdot \,  {\tilde K}(k)]\}, 
\end{eqnarray}
and for  $ \, T \, < \, T_c \, $  ($k_< \, = \, s_v^{-1} \, s_h^{-1}$):
\begin{eqnarray}
\hspace{-0.97in}&&  \quad \quad  \quad
a_{<0}\,\,  =\,\, \,  
{\sqrt{1+\nu k_<}} \cdot \, \{(1+k_</\nu) \cdot \, 
{\tilde \Pi}(-\nu k_< ,k_<) \, \, -(k_</\nu) \cdot \, {\tilde K}(k_<)\}, 
\nonumber\\
\hspace{-0.97in}&&  \quad \quad \ \quad
a_{<\mp1}\,\,  = \, \, \,  \, 
\frac{1}{(1+k_<\nu)^{1/2} \, \pm \, 1} 
\cdot \,  \{{\tilde E}(k_<) \,\, +(k_<^2-1) \cdot \, {\tilde K}(k_<)
\\
\hspace{-0.97in}&&  \, \quad \quad \quad  \quad \, \, 
\quad \pm \, (k_</\nu) \cdot \, 
(1+\nu/k_<)^{1/2} \cdot \, [(1+\nu k_<) 
\cdot \, {\tilde\Pi}(-\nu k_< , \,k_<)
 \, \, -{\tilde K}(k_<)]\}.
\nonumber
\end{eqnarray}
More generally, the $\, a_{n}$ 
(for $ \, T \, > \, T_c$ or $ \, T \, < \, T_c$)
are, also, all of the form
\begin{eqnarray}
\hspace{-0.97in}&&  \quad \quad  \quad \quad  \quad\quad
a_{n} \,\,  = \, \, \,  \, \alpha_n \cdot \, {\tilde K}
 \, \,+ \, \beta_n \cdot \, {\tilde E} 
\,\, + \, \gamma_n \cdot \, {\tilde \Pi}, 
\end{eqnarray}
where the coefficients $\, \alpha_n$, $\, \beta_n$, 
$\, \gamma_n$, are algebraic expressions of $\, k$ 
and $\, \nu$ (and, in fact, rational expressions in 
$  \, \,z_v= \, \, \tanh \beta E_v \,\, $ and 
$ \,\, z_h= \, \tanh \beta E_h$).

The general correlation function $ \, C(M,N)$ is 
given in what is called a bordered Toeplitz determinant
which, for $ \, N \, > \, M$, is an 
$ \, N \, \times N$  determinant whose
matrix elements in the first $ \, M$ columns are the $\, a_n$ of the
diagonal correlations. These matrix elements contain only the complete
elliptic integrals $ \, {\tilde K}$ and $\, {\tilde E}$. The matrix 
elements of the {\em remaining columns} are the matrix elements 
of the row correlations which are
{\em all linear combinations of the three types of elliptic integrals},
namely for $\, T \,> \,  T_c$, $ \, {\tilde K}(k)$, $\, {\tilde E}(k)$ 
and $\, {\tilde\Pi}(-k\nu,k)$. This generalizes the computation 
of the next to diagonal correlation of Au-Yang and Perk~\cite{ayp},
 but the details seem not to be in the literature 
(even in~\cite{witte2015}).    

\subsubsection{Homogeneous polynomials \\}
\label{Homogen}

\vskip .1cm

From these determinantal representations, and from
the fact that the $\, a_n$ entries 
are linear combinations of all three types of elliptic integrals,
 we conclude that the
correlation functions $ \, C(M,N)$, for instance  for 
$\, T \,> \,  T_c$, are, for $ \, N \, > \, M$, 
{\em homogeneous polynomials} 
of degree $ \, N$  in $ \, {\tilde K}(k)$, $\, {\tilde E}(k)$ 
and $\, {\tilde\Pi}(-k\nu,k)$, 
which contains all powers of $ \, {\tilde E}$ 
and $ \, {\tilde K}$ but only powers of $ \, \tilde \Pi$ of 
orders less than or equal to $ \, N \, -M$.
For  $ \, N \, < \, M$ they are homogeneous polynomials 
of degree $ \, N$  in $ \, {\tilde K}(k)$, $\, {\tilde E}(k)$ and 
$\, {\tilde\Pi}(-k/\nu,k)$, where the order in $\, {\tilde\Pi}(-k/\nu,k)$
is  less than or equal to $ \, M \, -N$.
For general values of $\, M$ and $\, N$, getting the exact expressions 
of the matrix elements of  {\em all the  remaining columns}
as {\em linear combinations of the three types of elliptic integrals},
is quite difficult. This determinantal approach is not an efficient 
and practical approach to get exact expressions of correlation 
functions that are not diagonal, next-to-diagonal, or row correlation 
functions\footnote[1]{The $\, a_n$ entries for the row correlation
verify a {\em five term linear recursion}, see equation (116) 
in~\cite{witte2015}.}.

\vskip .1cm

{\bf Remark 1:} The fact that the Ising model 
correlation functions can 
be expressed as sums of products
 of three complete elliptic integrals is 
not specific of two point correlation 
functions. Recalling~\cite{Groeneveld1,Groeneveld2}
one can reduce, for even integer $\, n$, 
any $\, n$-point correlation
function of the square lattice Ising model (or more 
general planar lattices) to sum of products
of two points correlation functions. Consequently, 
any even number correlation functions can also
be expressed as sums of products 
of three\footnote[2]{Or homogeneous polynomials
of four complete elliptic integrals, 
${\tilde K}(k)$, $\, {\tilde E}(k)$, 
$\, {\tilde\Pi}(-k\nu,k)$
and $\, {\tilde\Pi}(-k/\nu,k)$, because  $ \, N \, > \, M$,
as well as  $ \, N \, < \, M$, two-point correlation
functions both occur in the decomposition of a general
$\, n$-point correlation function.} complete 
elliptic integrals (${\tilde K}(k)$, $\, {\tilde E}(k)$ 
and $\, {\tilde\Pi}(-k\nu,k)$).

\vskip .1cm

{\bf Remark 2:}
The fact that the Ising model correlation functions can 
be expressed as sums of products
 of three complete elliptic integrals 
is reminiscent of the results of Boos et al.~\cite{Boos} 
where it was shown that
some correlation functions\footnote[8]{The correlation 
functions in~\cite{Boos} are ground state averages
of the product of spin operators on consecutive
columns on a same row of the lattice, to be compared
with the traditional definition of correlations functions 
of lattice spin models (see, for instance, 
equation (14) in~\cite{mpw}).}, associated 
with the eight-vertex model, can be expressed in terms 
of sum of products of three transcendental 
functions\footnote[5]{These three transcendental functions 
are three log-derivatives of a function expressed in terms 
of theta functions and the elliptic gamma function 
(see equation (2.32) in~\cite{Boos}). Seeing our results
(homogeneous polynomials
and more generally, sums of products of three complete 
elliptic integrals), as a straight subcase of the results
in~\cite{Boos} for the eight-vertex model is
not obvious (see the remarks in the first footnote
of~\cite{Boos}).}.

\subsection{Quadratic recursion relations}
\label{Quadra}

The efficient way to compute the exact expressions of the 
anisotropic correlation functions amounts, in fact, to using 
the {\em quadratic difference equations} obtained by McCoy 
and Wu~\cite{mwdiff1}-\cite{mwdiff2},
and by Perk~\cite{perk}.
These recursion relations relate the (high-temperature) 
correlation functions $ \, C(M,N)$ for $T \,> \, T_c$ to
the {\em dual correlation} $\, C_d(M,N)$ for $T \,> \, T_c$.
The {\em dual correlation function}  $\, C_d(M,N)$ 
{\em is defined as the low temperature correlation with the replacement}:
\begin{eqnarray}
\label{replacement}
\hspace{-0.97in}&&  \, \quad \quad \quad \quad \quad 
s_v \quad \longrightarrow \quad   {{1} \over { s_h}} 
\quad  \quad \quad \hbox{ and} \quad  \quad \quad \quad 
s_h \quad  \longrightarrow \quad  {{1} \over { s_v}}. 
\end{eqnarray}
These difference equations read
\begin{eqnarray}
\label{eqn1}
\hspace{-0.97in}&&  \, \quad \quad  \quad 
s^2_h \cdot \,[C_d(M,N)^2\, -C_d(M,N-1) \cdot \, C_d(M.N+1)]
\nonumber\\
\hspace{-0.97in}&& \quad  \,\quad  \quad \quad \quad \quad
+[C(M,N)^2\, -C(M-1,N) \cdot \, C(M+1,N)]
 \,\,\,  =\,\, \, \,0, 
\\
\label{eqn2} 
\hspace{-0.97in}&& \quad \quad \quad
 s^2_v \cdot \,[C_d(M,N)^2\, -C_d(M-1,N) \cdot \, C_d(M+1,N)]
\nonumber\\
\hspace{-0.97in}&& \quad \quad \quad  \quad \quad \quad
+[C(M,N)^2\, -C(M,N-1) \cdot \, C(M,N+1)]
\,\,\,   =\, \, \,\,0, 
\\
\label{eqn3}
\hspace{-0.97in}&& \quad \quad \quad
s_v s_h \cdot \,
[C_d(M,N) \cdot \, C_d(M+1,N+1)\, -C_d(M,N+1) \cdot \, (C_d(M+1,N)]
\nonumber\\
\hspace{-0.97in}&& \quad \quad \quad \quad \quad \quad
 =\,\,\,  C(M,N) \cdot \, C(M+1,N+1)\, -C(M,N+1) \cdot \, C(M+1,N), 
\end{eqnarray}
which hold for all $\,M$ and $\,N$, except 
$\,M=\,0,\,N =\,0$,
where we have:
\begin{eqnarray}
\label{eqn4}
\hspace{-0.97in}&& \quad \quad \quad 
 \quad \quad  \quad \,\, 
C(1,0) 
\,\,\, = \,\,\,\, (1+ s^2_h)^{1/2} \,\, -s_h \cdot \, C_d(0,1), 
\\
\label{eqn5}
\hspace{-0.97in}&& \quad \quad \quad 
 \quad \quad \quad \,\, 
C(0,1) \,\,\, = \,\,\,\,
(1+ s^2_v)^{1/2} \,\, - s_v \cdot \,C_d(1,0).
\end{eqnarray}
All the correlations may be computed from these quadratic 
relations using\footnote[2]{This system of (overdetermined) 
quadratic equations (\ref{eqn1}), (\ref{eqn2}),  (\ref{eqn3})
has many solutions, for 
instance a one-parameter family of solution 
$\, C(M,N, \lambda)$ corresponding to a 
$\, \lambda$-extension of the correlation functions, 
associated with other ``initial'' 
conditions for the quadratic recursions (see, for instance,
equations (65), (100) in~\cite{Mangazeev}). The $\, C(M,N)$ 
(here for $\, \lambda \, = \, 1$) are deduced, 
in a unique way, from these quadratic recursions
with these initial conditions: the initial conditions 
cannot be ``arbitrary''.} the diagonal correlation 
functions $\, C(N,N)$ and the first nearest 
neighbor correlations $\, C(0,0)\, = \, 1$, $\, C_d(0,0)\, = \, 1$, 
$\, C(0,1)$ and $ \, C(1,0)$ as input. Once one gets the 
expressions of the $\, C(M, \, N)$ and the {\em dual correlations} 
$\, C_d(M, \, N)$ from these {\em overdetermined}\footnote[1]{They 
can be viewed as discrete Painlev\'e lattice recursions.} set 
of quadratic relations, 
as homogeneous polynomials in $\, {\tilde E}(s_vs_h)$, 
$\, {\tilde K}(s_vs_h)$ and $\, {\tilde \Pi}(-s_h^2,s_vs_h)$, 
one can be confident in these exact expressions, the smallest 
miscalculation breaking immediately the rigid compatibility 
between these overdetermined set of quadratic relations.

Let us display the exact expressions of the first anisotropic 
correlation functions for $\, T\, >\, T_c$ where 
$\, k \, = \, \, s_vs_h$. We introduce the lighter notations:
\begin{eqnarray} 
\label{lighter}
\hspace{-0.97in}&& \quad \quad 
{\tilde E} \,\,  =\,\,\,  {\tilde E}(s_vs_h),
\quad \quad \quad 
{\tilde K} \,\,  = \,\, \, {\tilde K}(s_vs_h), 
\quad \quad \quad 
{\tilde \Pi}\,\,  = \,\, \, {\tilde \Pi}(-s_h^2,s_vs_h).
\end{eqnarray}
The first diagonal correlation functions read: 
\begin{eqnarray}
\label{Theandiag}
\hspace{-0.95in}&&  \, \,   
C(1, \, 1) \, \, = \, \, \, \,\, 
 {{{\tilde E}} \over {s_h \, s_v}} \, \,  \,
+ \, \, {{s_h^2 \, s_v^2 \, -1} \over {s_h \, s_v}} \cdot \, {\tilde K}, 
\\
\label{Theandiag2}
\hspace{-0.95in}&&  \, \, 
C(2, \,2) \,  \,=  \, \, \, {{1} \over {3}} \cdot  \, 
{{ 5 \, -\, s_h^2\, s_v^2} \over { s_h^2\, s_v^2}} \cdot \, {\tilde E}^2
\, \, + {{8} \over {3}} \cdot  \, 
{{s_h^2\, s_v^2 \, -1 } \over { s_h^2\, s_v^2}} 
\cdot \, {\tilde E} \, {\tilde K}
\, \, 
+ \,  {{(s_h^2\, s_v^2 \, -1)^2 } \over { s_h^2\, s_v^2}}
 \cdot \, {\tilde K}^2, 
\end{eqnarray}
The first row correlation functions
read: 
\begin{eqnarray}
\label{rowcorrel}
\hspace{-0.95in}&&  \, \, \quad \quad \quad \quad 
C(0, \, 1) \, \, = \, \, \, \,
 (s_v^2 \, +1)^{1/2} \cdot \, 
\Bigl( {{(s_h^2 \, +1)} \over { s_h}}
 \cdot \, {\tilde \Pi} \,  \,  -{{ {\tilde K}} \over {s_h}}   \Bigr), 
\end{eqnarray}
\begin{eqnarray}
\label{rowcorrel2}
\hspace{-0.95in}&&  \, \, 
C(0, \,2) \,  \,=  \, \, \, \,
{{s_h^2\,s_v^4 \, +s_v^4 \, +s_v^2 \, +1
} \over {s_h^2}} \cdot \, {\tilde K}^2 \,
\,\, \,-{{{\tilde E}^2} \over { s_h^2}} 
\, \, \, \, -2 \cdot \, {{(s_v^2 \, +1)^2 \cdot \, (s_h^2 \, +1) 
} \over { s_h^2 }} \cdot \, {\tilde K} \cdot \,  {\tilde \Pi}
\nonumber \\
\hspace{-0.95in}&& \quad  \quad \quad  \quad  \quad \quad  \quad 
\, + \,  {{(s_v^2 \, +1) \cdot \, (s_h^2 \, +1) 
\cdot \, (s_h^2 \, +s_v^2 \, +\, 2)
} \over { s_h^2 }} \cdot \,  {\tilde \Pi}^2, 
\end{eqnarray}
The first off-diagonal, off-row correlation function reads:
\begin{eqnarray}
\label{offcorrel}
\hspace{-0.95in}&&  \, \, \,  C(1, \, 2) \, \, \, = \, \, \, \,
 (s_v^2 \, +1)^{1/2} \cdot \, 
\Bigl( {{{\tilde E}^2} \over {s_h^2\,s_v  }} \,\,
 -\, {{s_h^2\,s_v^2 \, -1 } \over {s_h^2\,s_v }} 
\cdot \, {\tilde K}^2 \, \,  \, 
+ \, {{s_h^2\,s_v^2 \, +s_v^2 \, -2 } \over {s_h^2\,s_v }}
 \cdot \, {\tilde E} \, {\tilde K} 
\nonumber \\
\hspace{-0.95in}&& \, \, \,    \quad \quad \quad \quad
\, \, -\, {{ (s_h^2\, \, +1)  \, ( s_v^2\, \, -1) 
} \over {s_h^2\,s_v }} \cdot \, {\tilde E} \cdot \, {\tilde \Pi}
\, \, \,  \,  
+ {{ (s_h^2\, \, +1)  \, ( s_h^2\, s_v^2\, \, -1) 
} \over {s_h^2\,s_v }} \cdot \, {\tilde K} 
\cdot \, {\tilde \Pi} \Bigr).
 \end{eqnarray}
The corresponding dual correlation functions read respectively:
\begin{eqnarray}
\label{diagdual}
\hspace{-0.95in}&&  \, \,
C_d(1, \, 1) \, \, = \, \, \, {\tilde E}, 
\\
\label{diagdual2}
\hspace{-0.95in}&&  \, \,  
C_d(2, \,2) \,  \,=  \, \, 
{{1} \over {3}} \cdot  \, 
{{ 5 \, s_h^2\, s_v^2 \, -\, 1} \over { s_h^2\, s_v^2}} 
\cdot \, {\tilde E}^2
\, + {{2} \over {3}} \cdot  \, 
{{(s_h^2\, s_v^2 \, -1)^2 } \over { s_h^2\, s_v^2}} 
\cdot \, {\tilde E} \, {\tilde K}
\, - \,{{1} \over {3}} \cdot  \,  
 {{(s_h^2\, s_v^2 \, -1)^2 } \over { s_h^2\, s_v^2}} 
\cdot \, {\tilde K}^2,
\nonumber 
\end{eqnarray}
\begin{eqnarray}
\label{dualrowcorrel}
\hspace{-0.95in}&& \quad C_d(0, \, 1) \, \, = \, \, \,
(s_h^2 \, +1)^{1/2} \cdot \, \Bigl(
(s_v^2 \, +1) \cdot \, {\tilde \Pi}  \, 
 - s_v^2 \cdot \,  {\tilde K} \Bigr),
 \\
\hspace{-0.95in}&& \quad 
\label{dualrowcorrel2}
C_d(0, \,2) \,  \,=  \, \,  \,  \, \,  
{{s_h^2\,s_v^4 \, +2\, s_h^2\,s_v^2 \, +s_v^4 \, +s_v^2 \, -1
} \over {s_h^2}} \cdot \, {\tilde K}^2  \,  \,  \,
-2 \cdot \, {{s_h^2\,s_v^2 \, -1 } \over {s_h^2 }}
 \cdot \, {\tilde E} \, {\tilde K}
\, \, \,  \,  -{{{\tilde E}^2} \over { s_h^2}} 
\nonumber  \\
\hspace{-0.95in}&& \, \, \,   
\quad \quad \quad  \quad \quad \quad \quad 
-2 \cdot \, {{s_v^2  \cdot \, (s_h^2 \, +1)^2 \cdot \, (s_v^2 \, +1) 
} \over { s_h^2 }} \cdot \, {\tilde K} \cdot \, {\tilde \Pi}
\nonumber \\
\hspace{-0.95in}&& \, \, \,    
\quad \quad \quad  \quad \quad \quad \quad 
\, + \, 
 {{(s_v^2 \, +1) \cdot \, (s_h^2 \, +1) \cdot \, 
(s_h^2 \, +s_v^2 \, +\, 2\, s_h^2\,s_v^2)
} \over { s_h^2 }} \cdot \, {\tilde \Pi}^2,  
\end{eqnarray}
\begin{eqnarray}
\label{dualoffcorrel}
\hspace{-0.95in}&&  \, \, \,  C_d(1, \, 2) 
\, \,\, = \, \, \,\,
 (s_h^2 \, +1)^{1/2} \cdot \, \Bigl(
 \, s_v^2  \cdot \, {{ s_h^2\,s_v^2 \, -1 } \over {s_h^2 }} 
\cdot \, {\tilde K}^2 \, \,  \, 
+ \, {{s_v^2\, -1 } \over {s_h^2 }}
 \cdot \, {\tilde E} \, {\tilde K} \, \,  
\, \, + {{E^2 } \over {s_h^2 }} 
\nonumber \\
\hspace{-0.95in}&& \, \, \,    \quad \quad \quad \quad
\, \, +\, {{ (s_h^2\, \, -1)  \, ( s_v^2\, \, +1) } \over {s_h^2 }}
 \cdot \, {\tilde E} \cdot \, {\tilde \Pi}
\, \, \,  \,   \,  
- {{ (s_v^2\, \, +1)  \, ( s_h^2\, s_v^2\, \, -1) } \over {s_h^2 }} 
\cdot \, {\tilde K} \cdot \, {\tilde \Pi} \Bigr). 
 \end{eqnarray}

Recalling that the {\em dual correlation functions} 
$\, C_d(M,N)$  are defined as the low temperature 
correlation with the replacement (\ref{replacement}),  
one immediately deduces the corresponding exact 
expressions of the low-temperature correlation 
functions from the previous exact expressions 
(\ref{diagdual}),  (\ref{dualrowcorrel}), 
(\ref{dualrowcorrel2}), (\ref{dualoffcorrel}), 
 of the dual correlation functions.

Introducing some (low-temperature) notations:
\begin{eqnarray} 
\hspace{-0.97in}&& \quad 
{\tilde E}_< \,\, =\,\,\, 
{\tilde E}( {{1} \over {s_v\, s_h }}),
\quad \quad \quad 
{\tilde K}_< \,\,  = \,\, \, 
{\tilde K}({{1} \over {s_v\, s_h }}), 
\quad \quad \quad 
{\tilde \Pi}_< \,\,  = \,\, \, 
{\tilde \Pi}(- {{1} \over {s_v^2 }}, \, {{1} \over {s_v\, s_h }}).
\end{eqnarray}
the corresponding low-temperature correlation functions are 
immediately deduced. One gets for instance:
\begin{eqnarray}
\hspace{-0.97in}&& \quad 
C_<(1,1)\, \, =\,\, \,  {\tilde E}_< , 
\label{diag1}
\end{eqnarray}
\begin{eqnarray}
\label{aniso3}
\hspace{-0.97in}&& \quad \quad \quad 
C_<(0,1)\,\,  = \,\,\,\, (1+s_v^{-2})^{1/2} 
\cdot \, \{(1+s^{-2}_h) \cdot \, 
{\tilde \Pi}_< \, \,\, -s_h^{-2} \cdot \, {\tilde K}_<\}, 
\end{eqnarray}
\begin{eqnarray}
\label{aniso4}
\hspace{-0.97in}&& \quad \quad \quad 
C_<(0,2)\,\,  = \,
 \, \,\, (1+s_h^{-2}) \cdot \, (1+s_v^{-2}) \cdot \,
(2s_v^{-2}s_h^{-2} \, +s_v^{-2} \, +s_h^{-2}) 
\cdot \,  s_v^2 \cdot \,  {\tilde \Pi}_<^2 
\nonumber\\
\hspace{-0.97in}&& \quad \quad
 -2 \cdot \, s_h^{-2} \cdot \, (s_h^{-2}+1) \cdot \, 
(s_v^{-2}+1)^2 \, s_v^2 \cdot \, {\tilde K}_<  
\cdot \, {\tilde \Pi}_< \,\,  
\, -2 \cdot \, (s_v^{-2}s_h^{-2}-1)\, s_v^2 
\cdot \, {\tilde E}_<  \cdot \,  {\tilde K}_<
\nonumber\\
\hspace{-0.97in}&& \quad \quad \quad
+(s_v^{-2}s_h^{-2}\, +s_v^{-2}s_h^{-2}\, +s_h^{-4} \, +s_v^{-2}-1)
 \cdot \, s_v^2 \cdot \, {\tilde K}_<^2 
\, \,\,\, \, -s_v^2 \cdot \, {\tilde E}_<^2.
\end{eqnarray}
Of course one deduces the exact 
expressions for the correlation functions 
$\, C(N, \, M)$ from the exact expressions for the 
correlation functions $\, \,C(M, \, N)\,$ permuting 
$\, s_h \leftrightarrow  s_v$ in the previous expressions. 

\vskip .1cm 

{\bf Remark:} The two-point correlation functions $\, C(M,N)$
and $\, C_d(M,N)$ (with $M \, < \, N$) are 
homogeneous polynomials in $\, {\tilde E}$, $\, {\tilde K}$
and $\, {\tilde \Pi}$, the coefficients of the 
corresponding monomials
being rational functions of $\,s_v$ and $\, s_h$ when 
$\, N\, -\, M$ is even, and rational functions, 
up to an overall $\, (1+s_v^{-2})^{1/2}$ factor, when 
$\, N\, -\, M$ is odd. In fact for $\, C(M,N)$,
the rational functions are very simple rational 
functions: a polynomial expression of $\,s_v$ and $\, s_h$,
 divided by a denominator of the form $\,s_v^M \, s_h^N$
(see (\ref{Theandiag}), (\ref{Theandiag2}), (\ref{rowcorrel}), 
(\ref{rowcorrel2}), (\ref{offcorrel})).

\vskip .1cm 

\subsubsection{Diagonal reduction.\\}
\label{diagreduc}

\vskip .1cm 
We remarked, in subsection (\ref{determin}), that the 
diagonal correlation functions $\, C(N, \, N)$ can be 
seen to correspond to the $\, \alpha_1\, = \, \, 0$ limit 
of the anisotropic row correlation functions 
(see (\ref{dn3})). The condition $\, \alpha_1\, = \, \, 0$ 
corresponds to a $\, s_h$ small, $\, s_v$ large limit:
\begin{eqnarray}
\label{limiteps}
\hspace{-0.97in}&&  
\, s_h \, = \, \, s \cdot \, \lambda, \quad  \,   
\, s_v \, = \, \, {{s} \over { \, \lambda}}, 
\quad  \,   
\lambda \, \, \longrightarrow \, \, 0, 
\quad  \,   k \, = \, \, s_h\, s_v 
\, = \, \, s^2 \, = \, \, {{1} \over {\alpha_2}},
 \quad  \,  \hbox{ finite}. 
\end{eqnarray}
In this limit (\ref{limiteps}), 
$\,  {\tilde \Pi}\,\,  = \,\, \, {\tilde \Pi}(-s_h^2,s_vs_h)$,
the complete elliptic integral 
of the third kind (\ref{lighter}) reads:
\begin{eqnarray}
\label{limitepsPi}
\hspace{-0.97in}&& \quad  \quad   \, \,
 {\tilde \Pi}\,
\,\,  = \,\, \, \,  {\tilde K} \, \,\,
+ \, {{{\tilde E} \, -{\tilde K} } \over { s^2}}
 \cdot \,  \lambda^2 \,
\,+ \, {{(s^4+2) \cdot \, {\tilde K} \, -2 
\cdot \, (s^4+1)  \cdot \, {\tilde E} } \over {3 \, s^4 }} 
 \cdot \,  \lambda^4 
\nonumber \\
\hspace{-0.97in}&& \quad  \quad  \quad  \quad  \quad  
\, + \, {{ (8\,s^8\, +7\,s^4 \, +8) \cdot \, {\tilde E} 
\, \,\,  -  (4\, s^8 \, +3\, s^4 \, +8)
\cdot \, {\tilde K}} \over { 15 \, s^6 }}
 \cdot \,  \lambda^6 
\,\, \,\, + \, \cdots  
\end{eqnarray}
and one verifies that, in this limit, $\, C(0, \, 1)$ given by 
(\ref{rowcorrel}) {\em actually reduces to}  $\, C(1, \, 1)$ 
given by (\ref{Theandiag}) (with $\, s_h\, s_v \, = \, s^2$), 
that $\, C(0, \, 2)$ given by (\ref{rowcorrel2}) 
{\em actually reduces to}  $\, C(2, \, 2)$ given by 
(\ref{Theandiag2}).

Note that this verification of the reduction of the  
anisotropic correlations $ \,C(0,N)$ to 
the diagonal correlations $ \,C(N,N)$ 
{\em requires more and more terms} 
in the $\, \lambda$-expansion (\ref{limiteps}): 
up to $\, \lambda^{2  N}$ to check
the $\,\, C(0, \, N)\, \rightarrow \,\, C(N, \, N)\,$ 
limit\footnote[1]{In fact, more generally in the limit 
(\ref{limiteps}), one has  
$\,\, C(M, \, N)\, \rightarrow \,\, C(N, \, N)\,$  
 for all $M\, < \, N$ (not just $M\,=  \, 0$).}.

\vskip .2cm 

\section{Duality}
\label{duality}

One can easily verify that the exact expressions 
(\ref{diagdual}),  (\ref{dualrowcorrel}), 
(\ref{dualrowcorrel2}), (\ref{dualoffcorrel}) 
of some dual correlations  $\, C_d(M, \, N)$ 
{\em can be obtained from the exact expressions} (\ref{Theandiag}),
(\ref{rowcorrel}), (\ref{rowcorrel2}), (\ref{offcorrel})
of the corresponding (high temperature) correlations 
$\, C(M, \, N)$, when one performs a very simple transformation 
on the +
(high-temperature) $\, s_h$ and $\, s_v$ variables, 
and {\em on the complete elliptic integrals of the first, 
second and third kind}:
\begin{eqnarray}
\label{Kramers3action}
\hspace{-0.95in}&&  
\quad   
C_d(M, \, N)(s_h, \,\, s_v, \,\, {\tilde E},
 \,\, {\tilde K}, \, \, {\tilde \Pi})
\\
\hspace{-0.95in}&&  \quad  \quad \quad  \quad  
\,  \,=  \,\, \, \, 
C(M, \, N)\Bigl( {{1} \over {s_v}},
 \,\,\,  {{1} \over {s_h}},\,  \,\,   \,
{{{\tilde E}} \over {s_h \, s_v}} \, 
+ \, \, {{s_h^2 \, s_v^2 \, -1} \over {s_h \, s_v}}
 \cdot \, {\tilde K}, \, 
\,\, \,\,  s_h \, s_v \cdot \, {\tilde K}, \, \,\,  \,
 s_h \, s_v \cdot \, {\tilde \Pi}\Bigr).
\nonumber 
\end{eqnarray}
This result actually generalizes Ghosh's result~\cite{ghosh} 
from the isotropic to the {\em anisotropic} case.  Ghosh's 
result~\cite{ghosh} gave a representation of the duality 
{\em only on the complete elliptic integral of the first and second kind} 
(see also equations (57) in~\cite{Chan}).

We sketch in \ref{Wannier} a proof that the {\em involutive} 
transformation (\ref{Kramers3action}),
can actually be seen as a representation of the 
{\em Kramers-Wannier duality}
on the complete elliptic integrals of the first, 
second {\em and third kind}. 

\vskip .2cm 

Because the matrix elements in the 
determinantal expressions for the correlation functions are all 
expressible as linear combinations of $\, {\tilde K}$, $\, {\tilde E}$
and  $\, {\tilde \Pi}$,
this representation (\ref{Kramers3action}) of the
duality on the complete elliptic integrals 
of the first, second and third kind {\em holds for all
correlations} $ \, C(M,N)$.
 
\vskip .1cm 

\subsection{Isotropic limit}
\label{isotropic}

The correlations in the isotropic limit have been extensively 
studied by Ghosh and 
Shrock~\cite{ShrockGhosh,ShrockGhosh2,ShrockGhosh3}.
In the isotropic limit the fundamental anisotropic segregation 
between $\, M >  \, N$ and $ \, M >  \, N$,  no longer exists, 
one has $\, C(M, \, N) = \,  C(N, \, M)$, 
$\, s_h \, = \, s_v \, = \, s$. A consequence of identity 
(\ref{thirdident}) is that the complete elliptic integrals of the  
third kind actually reduce to the complete elliptic integral
of the first kind $\, K$ (see (\ref{thirdident})): 
\begin{eqnarray}
\label{P1fromK}
\hspace{-0.95in}&&  \quad  \quad  \quad  \quad  \quad 
 \quad  \quad  \quad  \,  
{\tilde \Pi} \,   \, \, = \, \, \, \,  \,  \, 
 {{1} \over {2}} \cdot \, {\tilde K} \, \,\, \,  \, 
 - {{1} \over {2}} \cdot \, {{1} \over {1\, +s^2}},
\end{eqnarray}
and consequently, all the previous exact expressions 
of the correlation functions $\, C(M, \, N)$ and dual 
correlation functions $\, C_d(M, \, N)$, are expressed 
as non-homogeneous polynomials in only 
$\, {\tilde E}$ and $\, {\tilde K}$.
In the isotropic limit, the previous homogeneity property, 
is now ``hidden'' in the exact expressions of the 
correlation functions expressed in terms of polynomials 
in only $\, {\tilde E}$ and $\, {\tilde K}$. 
For instance $\, C(0, \, 2)$ and $\, C_d(0, \, 2)$ 
read 
\begin{eqnarray}
\label{C02iso}
\hspace{-0.95in}&& \quad    \quad 
C(0, \, 2) \,  \,=  \, \, \,  \, 
{{ (s^2+1) \, (s^2-1)^2} \over { 2\, s^2}} \cdot {\tilde K}^2
 \, \, \, \, - \, \,  {{{\tilde E}^2} \over {s^2}} \,\,\,  \,  
+ \, \,  {{s^2 \, +1 } \over {2\, s^2 }},
\\
\hspace{-0.95in}&& \quad    \quad 
C_d(0, \, 2) \,  \, =  \, \, \, \,
 -\, {{ (s^2+1) \,  (s^2+2) \, (s^2-1)^2} \over { 
2\, s^2}} \cdot {\tilde K}^2 \,  \, \, 
\, - \, 2 \cdot \, 
{{ (s^2+1) \, (s^2-1)} \over {  s^2}} 
\cdot \, {\tilde E} \, {\tilde K} 
\nonumber \\
\label{Cd02iso}
\hspace{-0.95in}&& \quad    \quad   \quad   
\quad \quad   \quad    \quad  \quad \, 
- \, \,  {{{\tilde E}^2} \over {s^2}} \, \, \,\,  \, 
+ \, \,  {{s^2 \, +1 } \over {2}}.
\end{eqnarray}
and $\, C(1, \, 2)$ and $\, C_d(1, \, 2)$ read respectively:
\begin{eqnarray}
\label{C12iso}
\hspace{-0.95in}&&   
C(1, \, 2) \,  \,=  \, \, \,  (1\, +s^2)^{1/2} \cdot \, 
\Bigl(
 {{(s^2+1)\, (s^2-1)^2 } \over {2 \, s^3}} 
\cdot \, {\tilde K}^2 \, \, 
    + \,  {{ E^2 } \over {s^3 }} \, \, 
  + \,  {{(s^2+3)\, (s^2-1)} \over {2 \, s^3 }} 
\cdot \,     {\tilde E}  \, {\tilde K}   
\nonumber \\
\hspace{-0.95in}&&   \quad   \quad   \quad \quad   \quad       
  + \,  {{ (s^2+1)\, (s^2-1)} \over {2 \, s^3 }}
 \cdot \,  {\tilde K}   \, \, 
  - \,  {{ s^2-1} \over {s^3}} \cdot \,  {\tilde E} \Bigr),  
\end{eqnarray}
and
\begin{eqnarray}
\label{Cd12iso}
\hspace{-0.95in}&&     
C_d(1, \, 2) \,  \,=  \, \, \,
  (1\, +s^2)^{1/2} \cdot \, \Bigl(
 {{(s^2+1)\, (s^2-1)^2 } \over {2 \, s^2}} 
\cdot \, {\tilde K}^2 \, \, 
    + \,  {{ {\tilde E}^2 } \over {s^2 }} \, \, 
  + \,  {{(s^2+3)\, (s^2-1)} \over {2 \, s^2 }}
 \cdot \,     {\tilde E}  \, {\tilde K}   
\nonumber \\
\hspace{-0.95in}&&  \quad   \quad    
\quad   \quad   \quad     \quad      
  - \,  {{ (s^2+1)\, (s^2-1)} \over {2 \, s^2 }} 
\cdot \,  {\tilde K}   \, \, 
  + \,  {{ s^2-1} \over {s^2}} \cdot \,  {\tilde E} \Bigr). 
\end{eqnarray}

As it should, one does verify, on these  {\em non-homogeneous} 
exact expressions in only $\, {\tilde E}$ and $\, {\tilde K}$, 
that {\em all the dual correlation functions}  $\, C_d(M, \, N)$ 
{\em can also be deduced from the correlation functions}  
$\, C(M, \, N)$, using the representation of the Kramers-Wannier 
duality (\ref{Kramers}), rewritten in $\, s$:
\begin{eqnarray}
\label{Cd12isodual}
\hspace{-0.95in}&&  \quad    \quad   
C_d(M, \, N)(s, \, {\tilde E}, \, {\tilde K}) 
\,  \, \, \, 
=  \, \,\, \, \, C(M, \, N)\Bigl( {{1} \over {s}}, \, \, \,
{{{\tilde E}} \over {s^2}} \, 
+ \, \, {{s^4 \, -1} \over {s^2}} 
\cdot \, {\tilde K}, \,\, \, s^2 \cdot \, {\tilde K}\Bigr),
\end{eqnarray}
which is nothing but the Ghosh's statement
 in~\cite{ghosh}. This can be easily checked on (\ref{C02iso}) 
and (\ref{Cd02iso}), or (\ref{C12iso}) and (\ref{Cd12iso}). 
The fact that, when $\, M \, -N$ is odd the exact expressions of 
the correlation functions  $\, C(M, \, N)$ 
and of the dual correlation functions  $\, C_d(M, \, N)$ 
are very close (which is obvious on  (\ref{C12iso}) 
and (\ref{Cd12iso})), had already been remarked in the  Shrock 
and Ghosh paper~\cite{ShrockGhosh}.
It corresponds to the simple identity valid {\em only  when} 
$\, M \, -N \, \, $  {\em is odd}:
\begin{eqnarray}
\label{simpleidentity}
\hspace{-0.95in}&&  \quad \quad \quad   \quad  \quad   
C_d(M, \, N)(s, \, E, \, K) 
\,  \, \, =  \, \, \, \, s \cdot \, C(M, \, N)(s, \, -E, \,\, -K),
\end{eqnarray}
which is easily checked on (\ref{C12iso}) and (\ref{Cd12iso})
(and of course, not on (\ref{C02iso}) and (\ref{Cd02iso})).

\vskip .2cm 

\section{Linear differential equations for $ \, C(M,\, N)$} 
\label{lindiff}

The diagonal correlation $ \,C(N,N)$ was shown in~\cite{pvf} 
to satisfy a linear differential equation of order 
$ \, N+1$. For the isotropic case it was also seen 
in~\cite{fvp} that, for odd $ \,N$,  the row 
correlation $ \,C(0,N)$ satisfies 
an equation of order $ \,(N+1)(N+2)/2$ and for $ \,N$ even 
of order $ \,(N+2)^2/4$. 

Here we will consider, in the anisotropic case,
 the general correlations $ \,C(M, \, N)$ 
 and show that they are solution of a linear (partial) 
differential operator of order 
\begin{eqnarray}
\label{order}
\hspace{-0.99in}&& \quad \quad \quad \quad 
 {{1} \over {2}} \cdot \, (N \, -M \, +1) \cdot \, (N \, +M \, +2)
\\
\hspace{-0.97in}&& \quad \quad \quad 
\quad \quad \quad  \quad \quad  \, = \, \, \,   
{{1} \over {2}} \cdot \, (N  \, +1) \cdot \, (N  \, +2)
 \,\, \,         
- \, {{1} \over {2}} \cdot \, M \cdot \, (M  \, +1).
\nonumber 
\end{eqnarray}
 
The three functions in (\ref{lighter}), 
$ \, {\tilde \Pi}  \, = \, \,  {\tilde \Pi}(-\nu \, k, \,k)$,  
$\,{\tilde K} \, = \, \, {\tilde K}(k)$, 
$\, {\tilde E}\, = \, \, {\tilde E}(k)$, and their   
derivatives, form a three-dimensional vector space:
\begin{eqnarray}
\label{eq1}
\hspace{-0.97in}&& \quad \quad \quad \quad 
\frac{\partial \, \, {\tilde \Pi}_w}{\partial k} 
\, \, = \, \, \, 
 - \, {{ \nu \, + \, k } \over {2 \, \nu \, k \cdot \, w }}  
 \cdot \, {\tilde K} 
\,\,\,
+  \,{{ \nu \, k^2 \, + \, 2 \, k \, + \, \nu } \over {
2 \, \nu \, k \cdot \, (1-k^2) \cdot \, w }} 
  \cdot \, {\tilde E}, 
 \\
\label{eq2}
\hspace{-0.97in}&& \quad \quad \quad \quad 
 \frac{\partial {\tilde K}}{\partial k}
\,\, = \, \,\,\, 
 \frac{{\tilde E}}{k\cdot \, (1-k^2)} \, \,
\, - \, \frac{1}{k}\cdot \, {\tilde K}, 
 \\
\label{eq3}
\hspace{-0.97in}&& \quad \quad \quad \quad 
 \frac{\partial {\tilde E}}{\partial k}
\,\, =\,\, \, \, 
\frac{1}{k} \cdot \, \Bigl({\tilde E} 
\, \, -{\tilde K}\Bigr),
\quad \quad \quad \quad \quad \quad
 \hbox{where:}
\end{eqnarray}
\begin{eqnarray}
\label{Introduce2w}
\hspace{-0.95in}&&  \, \, \quad \quad \quad \quad 
{\tilde \Pi}_w \, \, = \, \, \, w \cdot \, {\tilde \Pi}
\quad \quad \quad \hbox{with:} \quad \quad \quad 
w \, \,\,  = \, \, \, \, 
    \Bigl((1 \, +\nu \, k)\, (1\, +{{k} \over {\nu}})\Bigr)^{1/2}.
\end{eqnarray}
Do note that introducing $\, {\tilde \Pi}_w$,
the product  of $\, {\tilde \Pi}$ 
with $\, w$, instead of $\, {\tilde \Pi}$, gives 
in (\ref{eq1}) a partial derivative
with respect to $\, k$ which is a  linear combination of 
$\, {\tilde E}$ and $\, {\tilde K}$,  with 
{\em no complete elliptic integral 
of the third kind} $\,  {\tilde \Pi}$.

The general correlation function $\, C(M, \, N)$ 
(with $\, M \, < \, N$) being a homogeneous 
polynomial of degree $\, N$ in $\, {\tilde E}$, 
$\, {\tilde K}$ and $\, {\tilde \Pi}$, it is {\em also} 
a homogeneous polynomial of degree $\, N$ in 
$\, {\tilde E}$, $\, {\tilde K}$ and $\, {\tilde \Pi}_w$:
\begin{eqnarray}
\label{CON}
\hspace{-0.97in}&& \quad \quad \quad \quad
C(M,N)\,\,\, =\,\,\,\,
 \sum_{l=0}^{N-M}\sum_{{m,n=0}\atop{l+m+n=N}}\, \, 
f_{l,m,n}^{(0)}(\nu,\, k) \cdot \, 
{\tilde \Pi}_w^l \cdot \, {\tilde K}^m \cdot \,  {\tilde E}^n, 
\end{eqnarray}
It has $\,(N -M+1)(N+ M+2)/2$ monomials 
$\, {\tilde \Pi}_w^l \, {\tilde K}^m  \,  {\tilde E}^n$
in the sum. It follows from (\ref{eq1})-(\ref{eq3}) 
that all the $\,(N -M +1)(N +M +2)/2$ first partial 
derivatives with respect to $\, k$ of $\,C(M, \, N)$ 
{\em will also be homogeneous polynomials} in 
$\, {\tilde E}$, $\, {\tilde K}$ and $\, {\tilde \Pi}_w$, 
each derivation decreasing the 
degree in $\, {\tilde \Pi}_w$, because of (\ref{eq1}):
\begin{eqnarray}
\label{CONp}
\hspace{-0.99in}&& \quad \quad \quad  \quad 
{{\partial^p C(M, \, N)} \over {\partial k^p}} \, = \,\,
 \sum_{l=0}^{N-M}\sum_{{m,n=0}\atop{l+m+n=N}} 
f_{l,m,n}^{(p)}(\nu,\, k) \cdot \,
{\tilde \Pi}_w^l\cdot  \, {\tilde K}^m \cdot \,   {\tilde E}^n, 
 \\
\hspace{-0.99in}&& \quad \quad  \quad  \quad  \quad \,\,\,
\hbox{where:}  
\quad \quad \quad \quad \quad \, \,    
 p \, =\, \,   1, \, \, \cdots, \,\, 
 {{(N -M +1)(N +M+2)} \over {2}}.
\nonumber
\end{eqnarray}
If one considers the systems of (partial) differential 
equations (\ref{CON}) and (\ref{CONp}), one can obtain all 
the monomials 
$\, {\tilde \Pi}_w^l \, {\tilde K}^m  \,  {\tilde E}^n$ 
in terms of $\,C(M, \,N)$ and its 
$\,(N -M+1)(N + M +2)/2 \, -1$ first 
derivatives, and deduce, from (\ref{CONp}) with 
$\, p \, = \, \, \,(N \, -M \, +1)(N\, +M \, +2)/2$, 
that $\,C(M, \, N)$ is actually solution of a linear (partial) 
differential operator of order $\,(N -M +1)(N +M +2)/2$.

\vskip .1cm  

This order $\,\, (N -M +1)(N +M +2)/2 \, \, $ linear 
differential operator is not irreducible. It  
{\em actually factors in $\, N\, -M\, +1$ operators of 
decreasing orders} $\, N\,  \, +1$, 
$\, N$, ... $\, M\, +2$, $\, M\, +1$:
\begin{eqnarray}
\label{cano}
\hspace{-0.95in}&&  \quad \, \quad  
\quad \quad \quad \quad 
 M_{N+1} \cdot \,  M_{N} \cdot \, M_{N-1} 
\, \,  \, \cdots \,\,   \, \, M_{M\, +2} \cdot \, M_{M\,+1},  
\end{eqnarray}
It is shown in \ref{diffoperapp} for the row correlation 
functions $\, C(0, \, N)$, but this can be generalized 
easily for general $\, C(M, \, N)$ correlations:  
the $\, M_n$'s are 
{\em actually  homomorphic to the $\, (n-1)$-th symmetric 
power}\footnote[1]{For $\, n=1$,  $\, M_1$ is an order-one 
linear differential operator.}  
of the order-two operator $\, L_K$ 
annihilating $\,{\tilde K}$.

In the special isotropic case, where $\,\nu=\,1$, the 
operator (\ref{cano}) reduces to direct sums~\cite{pvf,fvp}.  

\vskip .2cm 

\section{Conclusion} 
\label{Conclusion}

We have shown that the anisotropic correlation functions 
$\, C(M, \, N)$ are {\em homogeneous polynomials}
of degree $\, N$ in the complete elliptic 
integrals $\, {\tilde E}$,  $\, {\tilde K}$
and  $\, {\tilde \Pi}$, when $\, M \, < \, N$. For 
 $\, M \, > \, N$  the anisotropic correlation functions 
$\, C(M, \, N)$ are homogeneous polynomials of degree $\, N$
in $\, {\tilde E}$,  $\, {\tilde K}$, the  complete 
elliptic integral of the third kind 
$\, {\tilde \Pi}(- \,\nu \, k, \, k)$ being now 
replaced by $\, {\tilde \Pi}(-\, k/\nu, \, k)$. This 
remarkable property is totally 
hidden\footnote[2]{And is even easy to miss 
(as we did in~\cite{pvf,fvp}),
even in the anisotropic case if one uses 
$\, {\tilde \Pi}(- \,\nu \, k, \, k)$ for $\, M \, > \, N$
or $\, {\tilde \Pi}(-\, k/\nu, \, k)$ for $\, M \, < \, N$.} 
in the isotropic case (see equations for instance 
(\ref{C12iso}),  (\ref{Cd12iso})). This is not the first case 
a nice property, an important symmetry requires a larger 
framework to be seen: the best example is certainly 
the (star-triangle) Yang-Baxter integrability of the Ising model 
which cannot be seen on the isotropic square model, but actually 
requires to consider the {\em anisotropic} model. 
We obtained the exact expressions of these homogeneous polynomial
expressions from the quadratic recursions
 (\ref{eqn1}), (\ref{eqn2}), (\ref{eqn3}) 
of section (\ref{Quadra}). The homogeneous polynomial
character of the solutions of such a remarkably rigid and 
overdetermined set of quadratic equations  is worth noticing 
from an {\em integrable lattice maps}~\cite{Viallet}, or 
{\em discrete Painlev\'e}~\cite{Kanki,Sakai} viewpoint.

We have shown that the linear differential operators annihilating 
the anisotropic row correlation functions $\, C(0, \, N)$, are 
operators of order $ \, (N+1)(N+2)/2$, with a remarkable canonical 
factorization in a product of operators 
{\em homomorphic to the successive $\, n$-th symmetric powers} 
of the operator annihilating $\, {\tilde K}$, the complete 
elliptic integral of the first kind, and similar results can be 
obtained for the general anisotropic correlation functions 
$\, C(M, \, N)$.

All the results given in this paper underline the fundamental
role of the complete elliptic integral of the {\em third kind} 
in order to have a clean, clear-cut description of the anisotropic 
Ising model~\cite{AnisotropicHolo}. Along this line, we have been 
able to extend Ghosh's representation of the Kramers-Wannier 
duality~\cite{ghosh} on the complete elliptic integral of the first 
and second kind, to a {\em representation of the Kramers-Wannier 
duality to  complete elliptic integral of the third kind}.
This involutive representation actually enables to {\em get the exact 
expressions of all the dual correlations $\, C_d(M, \, N)$ occurring
in the quadratic relations, from the exact expressions of 
the  correlations} $\, C(M, \, N)$. This representation of 
the Kramers-Wannier duality (\ref{Kramers3action}) can thus 
be seen as a {\em symmetry} of these overdetermined set of  
equations.

\vskip .2cm

\vskip .2cm

{\bf Acknowledgments:} 
One of us (JMM) would like to thank
 S. Boukraa, S. Hassani and J-A. Weil for very fruitful 
discussions  on the correlation functions of the anisotropic 
Ising model, and on differential systems.
One of us (BMM) would like to thank R.E. Shrock and N. Witte 
for many useful discussions about Ising correlations.

\vskip .5cm
\vskip .5cm

\appendix

\section{Proof of the identity on a complete elliptic integral of the third kind}
\label{proof}

As noted in section (\ref{nearest}) the identity (\ref{piidentity})  
can be seen as a consequence of addition
formulae on the Jacobi's Zeta function (see (\ref{Zeta})). We 
give here a perspective in terms of 
homomorphisms\footnote[1]{In the sense of the equivalence
of linear differential operators~\cite{vdPut} 
(corresponding to the 'Homomorphisms'
command in Maple's DEtools package).} 
of linear differential operators.

One can rewrite the identity (\ref{piidentity}) on the 
complete elliptic integral of the third kind
in the following form: 
\begin{eqnarray}
\label{piidentityappxy}
\hspace{-0.95in}&&  \, \,  \quad   \quad  \quad   \quad  
A(x, y) \cdot \, {\tilde\Pi}(x, y) \,\, \, 
+ \, \, B(x, y)  \cdot \, {\tilde\Pi}(R(x, \, y), y)
\, \,\, \, = \, \,  \,  \,  \, \, {\tilde K}(y), 
\end{eqnarray}
where:
\begin{eqnarray}
\label{piidentityapp2}
\hspace{-0.95in}&&  \, \quad   \quad \quad  \quad   \quad  
R(x, \, y) \, \, = \, \,  \,
4  \cdot \, y^2  \cdot \, 
\frac{ x  \cdot \, (x-1)  \cdot \,(x-y^2)}{(x^2 \, -y^2)^2},
\\
\hspace{-0.95in}&&  \, 
A(x, y)\,  = \, \,   4 \cdot \,  
{{ (1-x) \cdot \, (x -y^2) } \over { \rho}},
 \quad      \quad    
B(x, y) \,  = \, \,  
 {{ (x^2+y^2 \, -2 x) \cdot \, (x^2 +y^2-2 y^2 x)} \over {
(y^2 -x^2) \cdot \, \rho  }},
\nonumber \\
\hspace{-0.95in}&&  \, \quad    \quad     \quad 
\hbox{where:} \quad   \quad   \quad \quad\quad      \quad  
\rho  \, \, = \, \,  \, x^2 \,-2x\, -2 y^2\, x \, +3 y^2. 
\end{eqnarray}
The complete elliptic integral of the third kind 
$\, {\tilde\Pi}(x, y)$ is known to be (see~\cite{pvf}) 
solution of a third order linear differential operator, 
which is the product of an order-one linear differential 
operator and the square of one order-one linear 
differential operator:
\begin{eqnarray}
\label{piidentityappxyop}
\hspace{-0.95in}&&  \, 
{\cal L}_3 \, \, = \, \, \, 
\Bigl( {{\partial } \over {{\partial x}}}
 \, + \,  {\frac {2\,xy^2 -3\,{x}^{2} +2\,x -y^2}{ 
(x-1)  \cdot \, (y^2-x) \cdot \,  x}}\Bigr)^2
  \cdot \, \Bigl( {{\partial } \over {{\partial x}}}
  \, + \, \,  {{1} \over {2}} \,{\frac {y^2-{x}^{2}}{
 (x-1)  \cdot \, (y^2-x) \cdot \,  x}}\Bigr).
\end{eqnarray}
The complete elliptic integral of the third kind 
$\, {\tilde\Pi}(R(x, \, y), y)$ 
is solution of another third order linear differential
operator $\, {\cal L}_3^{(R)}$, which is $\, {\cal L}_3$ 
pullbacked by the {\em infinite order transformation}
 $\, x \, \rightarrow \, R(x, \, y)$. The operator 
$\, {\cal L}_3^{(R)}$ is, of course, also a product of three 
order-one operators. A remarkable identity like 
(\ref{piidentityappxy}) requires the two order-three linear 
differential operators  $\, {\cal L}_3$ and 
 $\, {\cal L}_3^{(R)}$ to be ``very closely related''. This 
is the case. These two order-three operators 
{\em are  actually homomorphic}\footnote[1]{This highly non-trivial 
situation of two homomorphic operators where the second one is the 
pullback of the first one by an infinite order rational function, 
has already been seen and described in~\cite{Renorm}.}.
The order-three operators  $\, {\cal M}_3$ 
and $\, {\cal N}_3$,  annihilating 
$\, A(x, y) \cdot \, {\tilde\Pi}(x, y)$ and 
$ \, B(x, y)  \cdot \, {\tilde\Pi}(R(x, \, y), y)$ are 
simply obtained from $\, {\cal L}_3$ and  $\, {\cal L}_3^{(R)}$ 
by conjugation by respectively $\, A(x,y)$ and 
$\, B(x,y)$, and are, of course, also homomorphic. 
The order-three operators  $\, {\cal M}_3$ 
and $\, {\cal N}_3$ are also product 
of order-one linear differential operators:
\begin{eqnarray}
\label{productofthree}
\hspace{-0.95in}&&  \quad  \quad 
{\cal M}_3 \,  \, = \, \,  \, 
\Bigl({{\partial } \over {{\partial x}}}\, 
+ {{\partial \ln(\rho_1)} \over {\partial x}} \Bigr)^2 \cdot \, 
 \Bigl( {{\partial } \over {{\partial x}}} \, 
+ {{1} \over{2}} \cdot \, {{\partial \ln(\rho_2)} \over {\partial x}} \Bigr), 
\\
\label{productofthree2}
\hspace{-0.95in}&&  \quad \quad 
{\cal N}_3 \,  \, = \, \,  \, 
\Bigl( {{\partial } \over {{\partial x}}} \, 
+ {{\partial \ln(\rho_3)} \over {\partial x}} \Bigr)
\cdot \, 
\Bigl( {{\partial } \over {{\partial x}}}\, 
+ {{\partial \ln(\rho)} \over {\partial x}} \Bigr)
\cdot \, 
\Bigl( {{\partial } \over {{\partial x}}} \, 
+ {{1} \over{2}} \cdot \, {{\partial \ln(\rho_2)} \over {\partial x}} \Bigr),
\end{eqnarray}
where:
\begin{eqnarray}
\label{rho1}
\hspace{-0.95in}&&   \quad  \quad  \quad \quad 
 \rho_1  \,  \, = \, \,  \, x \cdot \, \rho,
\quad  \quad  \quad \, \, 
\rho_2  \,  \, = \, \,  \, {{\rho^2} \over {(x-1)\cdot \, x \cdot \, (x \,-y^2)}},
\nonumber \\
 \hspace{-0.95in}&&  \, \quad \quad \quad \quad 
\rho_3  \,  \, = \, \,  \, 
{{(x-1)^2 \cdot \, (x\, -y^2) \cdot \, x^2 \cdot \, \rho } \over {
(x^2 \,+y^2 \, -2\,x\,y^2)\cdot \, (x^2 +y^2 \, -2\, x)\cdot \, (x^2 \, -y^2)}}. 
\end{eqnarray}
The lhs of identity (\ref{piidentityappxy}) 
is the sum of $\, A(x, y) \cdot \, {\tilde\Pi}(x, y)$ and 
$ \, B(x, y)  \cdot \, {\tilde\Pi}(R(x, \, y), y)$
and is thus solution of the LCLM\footnote[2]{The 
Least Common Left Multiple (LCLM) of linear
differential operators $\,L_1, \, \cdots, \, L_n$ 
is defined as the (minimal order) linear differential 
operator such that all solutions of 
$\,L_1, \, \cdots, \, L_n$
are solutions of this LCLM as well (see
for instance~\cite{vdPut}).} of $\, {\cal M}_3$ 
and $\, {\cal N}_3$. 
In the identity (\ref{piidentityappxy}), $\, y$ remains fixed, 
and thus the rhs of (\ref{piidentityappxy}),  
$\, {\tilde K}(y)$, is a constant 
with respect to the partial derivations in $\, x$. Proving identity 
(\ref{piidentityappxy}) requires the constant functions in $\, x$
to be solution of the previous LCLM. This is actually the case. This 
LCLM is an order-four operator, which is the product of four 
order-one operators, the order-one operator right divising 
this LCLM {\em being precisely} 
$\, \partial/\partial x$:
\begin{eqnarray}
\label{itisthecase4}
\hspace{-0.95in}&&  \,  \,  \, 
\quad 
{\cal M}_3 \oplus {\cal N}_3 
\, \,  \, \,   = 
\\
\hspace{-0.95in}&&  \,  \, \,  
\quad \quad 
\Bigl( {{\partial } \over {{\partial x}}} 
\, + {{\partial \ln(\rho_4 \cdot \, \rho)} \over {\partial x}} \Bigr)
\cdot \, 
\Bigl( {{\partial } \over {{\partial x}}}
\, + {{\partial \ln(\rho_5 \cdot \, \rho)} \over {\partial x}} \Bigr)
\cdot \, 
\Bigl( {{\partial } \over {{\partial x}}} 
\, + {{1} \over{2}} \cdot
 \, {{\partial \ln(\rho_6 \cdot \, \rho^4)} \over {\partial x}} \Bigr)
\cdot  \,  {{\partial } \over {{\partial x}}} ,
\nonumber 
\end{eqnarray}
where the exact expressions of  the three rational functions  
of the two variables $\, x$ and $\, y$, 
 namely $\, \rho_4$, $\, \rho_5$ and  $\, \rho_6$,
 is not important for this proof\footnote[1]{What matters is the 
rightdivision of the LCLM by
$\, \partial/\partial x$ which is not obvious. In contrast 
the rightdivision of the LCLM by 
$\,\, \partial/\partial x \, + \, 1/2 \cdot \,\partial \ln(\rho_2)/\partial x\,\,$
is obvious (see (\ref{productofthree}), (\ref{productofthree2})).
}. For 
instance $\, \rho_6$ reads: 
\begin{eqnarray}
\label{itisthecase5}
\hspace{-0.95in}&&  \, \quad \quad \quad  \quad \quad  \,  \,  \, 
\rho_6 \,  \, = \, \,  \, 
{{(x-1)\cdot \, x \cdot \, (x-y^2)} \over {
(4\,x\,y^4 \,+x^4 \, -6\,x^2\,y^2 \, -3\,y^4 \,+4\,x\, y^2 )^2 }}.
\end{eqnarray}

Finding the constant (in $\, x$) in the rhs of identity 
(\ref{piidentityappxy}) 
is easily done taking the $\, x=\, 0$ limit of the lhs of 
(\ref{piidentityappxy}). Since 
$\, {\tilde\Pi}(R(0, \, y), y) \, \, = \,\, \, {\tilde K}(y)$, 
the evaluation of the lhs of 
(\ref{piidentityappxy}) at $\, x \, = \, \, 0$ gives 
the rhs of identity (\ref{piidentityappxy}):
\begin{eqnarray}
\label{4over3}
\hspace{-0.95in}&&  \, \quad \quad \quad 
\quad \quad \quad \quad \quad 
{{ 4} \over {3}} \cdot \, {\tilde K}(y) \, \,\,
-{{ 1} \over {3}} \cdot \, {\tilde K}(y) \,
 \, \, \,= \,\, \,\,{\tilde K}(y).
\end{eqnarray}

\section{Representation of the Kramers-Wannier duality 
on the complete elliptic integrals}
\label{Wannier}

The Kramers-Wannier duality on the anisotropic model is well-known to
correspond to 
$\, (s_h, \, s_v) \, \rightarrow \, (1/s_v, \, 1/s_h)$, namely 
$\, k \,  \rightarrow \, 1/k$ when
$\, \nu \, = \, s_h/s_v$ {\em remains invariant}.

\subsection{Representation of the Kramers-Wannier duality 
on the complete elliptic integral of the third kind}
\label{third}

Let us consider the function 
$\, {\tilde \Pi} \, = \, \,  {\tilde \Pi}(- \, \nu \, k, \, k)$, 
pullbacked by 
$\, k \,  \rightarrow \, 1/k$,
$\, \nu$ {\em remaining invariant}:
\begin{eqnarray}
\label{pullback}
\hspace{-0.95in}&&  \, \, 
\quad \quad \quad \quad  \quad \quad  \quad \quad 
{\tilde \Pi}_{pull} \,\, = \, \, \, \,
{\tilde \Pi}\Bigl(-{{\nu } \over { k}}, \, {{1} \over { k}})\Bigr), 
\end{eqnarray}
$\, {\tilde \Pi}_{pull}$, given by (\ref{pullback}),
 is solution of an order-three linear differential 
operator $\, {\cal M}_3$ 
which also factorizes in the product 
of an order-two operator $\, {\cal M}_2$ and an  
order-one operator 
$\, {\cal M}_1$
where 
\begin{eqnarray}
\label{Introduce3M1}
\hspace{-0.95in}&&  \, \, \quad  \quad  \quad 
{\cal M}_1\, \, = \, \, \, \, 
{\tilde w} \cdot \,  \frac{\partial }{\partial k}
  \cdot \,   {{1} \over { {\tilde w} }} 
\, \, \, \,\,  = \, \, \,   \, \,
\frac{\partial }{\partial k} \,  \,\, 
- \, {{ (\nu^2+1) \cdot \, k \, +2\,\nu } \over {
 2 \cdot \, k \cdot \, (1\, + \, \nu \, k) \cdot \, (k \, + \, \nu )}}.
\end{eqnarray}
where $\, {\tilde w}$ is $\, w$  in (\ref{Introduce2w}) 
with $\,\,\, k \,  \longrightarrow \, 1/k$.

Now consider $\, k \cdot \, {\tilde \Pi}$, one finds 
that {\em it is also solution} 
of the {\em same order-three operator} $\, {\cal M}_3$.
The order-three operator $\, {\cal M}_3$ has 
this solution 
\begin{eqnarray}
\label{solukP1}
\hspace{-0.95in}&&  \, \, \quad  \quad  \quad  \quad 
Sol \,\,  = \, \,\, \, \, k \cdot \,  {\tilde \Pi}
\,\, \, \, = \, \, \,\,\, \,
 k \,\, \, \, \,  -{{ \nu} \over {2}}  \cdot  \,k^2 \, \, \, \, 
+{{3\,\nu^2\,+2} \over {8}} \cdot \, k^3 
\,\, \,\,   + \, \, \cdots
\end{eqnarray}
with, of course,  the solution of
the algebraic solution of  $\, {\cal M}_1$, namely 
\begin{eqnarray}
\label{soluS1}
\hspace{-0.95in}&&  \, \, \,  \quad \quad  \quad \quad 
S_0 \,\,  = \, \, \, \, 
{{ k} \over { 
(1\, + \, k/\nu)^{1/2} \cdot \, (1\, + \nu\, k)^{1/2}}} 
\\
\hspace{-0.95in}&&  \, \, \,  \quad \quad 
\quad \quad  \quad \quad 
\,\,  = \, \, \,\,\,\,\, 
k \, \,\, \, \, 
-{{1} \over {2}} \cdot  \, 
{\frac { ({\nu}^{2}+1) }{\nu}}  \cdot  \, k^2
\, \,\,  \,
+{{1} \over {8}} \cdot \, 
{\frac { (3\,{\nu}^{4}+2\,{\nu}^{2}+3) }{{\nu}^{2}}}
 \cdot  \,k^3 \,\,\,\,\,    + \, \,\,  \cdots
\nonumber 
\end{eqnarray}
together with the formal series solution
\begin{eqnarray}
\label{formal}
\hspace{-0.95in}&&  \, \, \quad  \quad  \quad   \quad  \quad  
Sol_f \,\,  = \, \, \, \, 
\Bigl(Sol \,  -{{S_0} \over {2}}\Bigr) \cdot \, \ln(k)
 \, \, \, + \, \, \,  {\cal A}_2, 
\end{eqnarray}
where $\, {\cal A}_2$ denotes an analytical series in $\, k$: 
\begin{eqnarray}
\label{formalA}
\hspace{-0.95in}&&  \, \, \quad  \quad  \,
{\cal A}_2 \, \,  = \, \,  \, \,\, 
{{1} \over {2}} \cdot \,{\frac {k}{{\nu}^{4}}} \,\,  \,\,   \, 
-{{3} \over {8}} \cdot \,{\frac {{k}^{2}}{{\nu}^{5}}} \, \, \, \, 
 + {{1} \over {32}} \cdot \, 
{\frac { (13\,{\nu}^{4}-4\,{\nu}^{2}+8)}{{\nu}^{6}}} \cdot \, k^3
 \, \, \,\,  \, + \, \,  \, \cdots 
\end{eqnarray}
$ {\tilde \Pi}_{pull}$, given by (\ref{pullback}), 
is {\em well-defined as a series 
expansion in} $\, 1/k$ (low temperature expansions for the 
Ising model), namely $\, k$ large. If one wants to understand 
it as a series expansion in $\, k$ ($k \simeq \, 0$), one knows, 
since it is solution of an order-three linear differential
operator $\, {\cal M}_3$, 
that it is a linear combination of $\, S_0$,  $\, Sol$, $\, Sol_f$ 
given by (\ref{solukP1}),  (\ref{soluS1}),   (\ref{formal}), but 
finding precisely which linear combination, amounts to performing 
some analytical continuation\footnote[1]{This is typically 
a calculation of the {\em connection matrix} from $\, k= \,\infty$ 
to $\, k=\, 0$, see for instance~\cite{ze-bo-ha-ma-05c}.} 
from $\, k= \,\infty$ to $\, k=\, 0$. Note that {\em only} 
the linear combinations of $\, S_0$ and  $\, Sol$  
{\em are analytical at} $\, k=\, 0$. 

\vskip .1cm  

\subsection{Representation of the Kramers-Wannier duality 
on the complete elliptic integral of the second kind}
\label{second}

Let us consider similar $\, k \, \rightarrow \, 1/k$
duality calculations on the two simpler complete elliptic 
integrals $\, {\tilde E}$ and $\, {\tilde K}$. One 
easily finds, similarly, that 
\begin{eqnarray}
\label{calE}
\hspace{-0.95in}&&  \, \, \quad  \quad \quad 
 \quad  \, \quad  \quad   \quad  
{\tilde E}_{pull} 
\,  \,= \, \, \, 
{\tilde E}\Bigl({{1} \over { k}}\Bigr), 
\end{eqnarray}
and 
\begin{eqnarray}
\label{Ed}
\hspace{-0.95in}&&  \, \, \quad  \quad \quad 
 \quad  \quad    \,  \, 
{\tilde E}_{d} \,\, = \, \, \, \,  \, 
{{{\tilde E}} \over {k}} 
\,\,  \,  + \, \, {{k^2 \, -1} \over {k}} \cdot \, {\tilde K}
\\
\hspace{-0.95in}&&  \, \,  
 \quad  \quad   \quad  \quad \quad  
\quad \quad  = \, \, \, \,  \, 
{\frac {k}{2}}  \, \, \, +{\frac {k^3}{16}}  
\, \, \, +{\frac {3}{128}} \cdot \,{k}^{5} \, \, \, 
+{\frac {25}{2048}} \cdot \,{k}^{7} \, \, 
\, \, + \, \, \cdots 
\end{eqnarray}
are solutions of the 
{\em same second-order linear differential operator}.
The other solution of this linear differential operator 
is the formal series
\begin{eqnarray}
\label{Edformal}
\hspace{-0.95in}&&  \, \, \quad   \quad  \quad 
\quad   \quad  \quad \quad  \quad  
E_f \, = \, \, \,  {\tilde E}_{d} \cdot \, \ln(k) \,  \,  \, \, + \, \,  \, E_a, 
\end{eqnarray}
where $\, E_a$ is a holonomic Laurent series  at $\, k\, = \, 0$:
\begin{eqnarray}
\label{Edformal2}
\hspace{-0.95in}&&  \, \, \quad  \quad   \quad   \quad  \quad   \quad  
E_a \,\, = \, \, \,\, \, 
-\, {{1} \over {k}}\,\,\, \,  + {{k} \over {8}} \,\,\, \, 
 +{\frac {3\,{k}^{3}}{32}} \,\, +{\frac {21\,{k}^{5}}{512}}
\,\,\,  +{\frac {1115\,{k}^{7}}{49152}} 
\,\,\,\,  \, + \, \, \, \cdots 
\end{eqnarray}
${\tilde E}_{pull}$, given by (\ref{calE}), is 
{\em well-defined as a series expansion in} 
$\, 1/k$ (low temperature expansions for the Ising model), 
namely $\, k$ large. If one wants to understand (\ref{calE})
as a series expansion in $\, k$ ($k \simeq \, 0$), one knows, since 
it is a solution of the same second-order linear differential operator,
as $\, {\tilde E}_{d}$, given by (\ref{Ed}), that (\ref{calE}) is a 
linear combination of $\,  {\tilde E}_{d}$ and of the formal 
series $\, E_f$ given by (\ref{Edformal}). Of course, 
{\em only} $\,  {\tilde E}_{d}$, given
 by (\ref{Ed}), {\em is analytic at} $\, k\, = \, \, 0$.

\subsection{Representation of the Kramers-Wannier duality 
on the complete elliptic integral of the second kind}
\label{first}

Finally, one also finds, similarly, that 
\begin{eqnarray}
\label{calK}
\hspace{-0.95in}&&  \, \, \quad  \quad  \, \quad
  \quad   \quad  \quad   \quad  
{\tilde K}_{pull}
\, = \, \, \, 
{\tilde K}\Bigl({{1} \over { k}}\Bigr), 
\end{eqnarray}
and 
\begin{eqnarray}
\label{Kd}
\hspace{-0.95in}&&  \, \, \quad  \quad  \, \quad
  \quad   \quad  \quad   \quad 
{\tilde K}_{d} \,\, = \, \, \, k \cdot \, {\tilde K}, 
\end{eqnarray}
are solutions of the 
{\em same second-order linear differential operator}.
The other solution of this linear differential operator 
is the formal series
\begin{eqnarray}
\label{Kdformal}
\hspace{-0.95in}&&  \, \, \quad  \quad  
\quad   \quad  \quad  \quad  
K_f \,\, = \, \,\, {\tilde K}_{d} \cdot \, \ln(k) 
\, \, \, + \, \,  \, K_a, 
\end{eqnarray}
where $\, K_a$ is a holonomic series analytic at $\, k\, = \, 0$:
\begin{eqnarray}
\label{Kdformal2}
\hspace{-0.95in}&&  \, \, \quad  \quad 
\quad   \quad   \quad   \quad  
K_a \,\, = \, \, \, \,
\,{{k^{3}} \over {4}}\,\, \,+{\frac {21\,{k}^{5}}{128}}\,\,\,
+{\frac {185\,{k}^{7}}{1536}}\,\, \,
+{\frac {18655\,{k}^{9}}{196608}}
\, \, \,\, \, + \, \cdots 
\nonumber 
\end{eqnarray}
${\tilde K}_{pull}$ given by (\ref{calK}), is 
{\em well-defined as a series expansion in} $\, 1/k$ 
(low temperature expansions for the Ising model), 
namely $\, k$ large. If one wants to understand (\ref{calK})
as a series expansion in $\, k$ ($k \simeq \, 0$), one knows, since 
it is a solution of the same second-order linear differential operator,
as $\, {\tilde K}_{d}$, given by (\ref{Kd}), that (\ref{calK}) is a 
linear combination of $\, {\tilde K}_{d}$ and of the formal series 
$\, K_f$ given by (\ref{Kdformal}). Of course, {\em only} 
$\, {\tilde K}_{d}$, given by (\ref{Kd}) {\em is analytic at} 
$\, k\, = \, \, 0$.

\subsection{Representation of the Kramers-Wannier duality}
\label{represe}

It had already been written in~\cite{ghosh} (see formulae (5), (6), (7) 
 in~\cite{ghosh}) that a representation of the Kramers-Wannier duality
on $\, {\tilde E}$ and $\, {\tilde K}$ actually 
reads\footnote[1]{See also Table 4, Transformations of complete 
elliptic integrals, page 319, 13.8, in Bateman.}: 
\begin{eqnarray}
\label{Kramers}
\hspace{-0.95in}&&  \, \, \, \, 
(k, \,\, {\tilde E}, \,\, {\tilde K}) \, 
\quad \longrightarrow \, \,  \quad 
\Bigl( {{1} \over {k}}, \, \, {\tilde E}_{d}, \, \, {\tilde K}_{d}\Bigr)
 \, \, = \, \,     \, 
\Bigl( {{1} \over {k}}, \, \, \,
{{{\tilde E}} \over {k}} \, + \, \, {{k^2 \, -1} \over {k}}
 \cdot \, {\tilde K}, \,\, \, k \cdot \, {\tilde K}\Bigr). 
\end{eqnarray}
Let us denote $\, {\tilde \Pi}_p$ the complete elliptic 
integral of the third kind $\, {\tilde \Pi}$,  where 
$\, s_h$ and $, s_v$ {\em have been permuted},
$\, s_h \, \leftrightarrow \, s_v$, namely 
 $\, {\tilde \Pi}(-k/\nu, \, k)$.

From the previous calculations (\ref{third}) 
on the complete elliptic 
integral of the third kind, it is natural to imagine that 
$\, {\tilde \Pi}$ and $\, {\tilde \Pi}_p$ transform as linear 
combinations of (\ref{solukP1}) and  (\ref{soluS1}), namely:
\begin{eqnarray}
\label{alphabeta}
\hspace{-0.95in}&&  \, \, \quad   \quad   \quad      \quad 
 ({\tilde \Pi}, \, \, {\tilde \Pi}_p) 
\, \quad    \longrightarrow \,\quad     \,  \, 
\Bigl(\alpha \cdot k \cdot \, {\tilde \Pi} \, + \,  \, \beta \cdot \, S_0, 
\,\,  \,  \alpha \cdot k \cdot \, {\tilde \Pi}_p 
\, + \,  \, \beta \cdot \, S_0\Bigr).
\end{eqnarray}
One has the same coefficients $\, \alpha$ and $\, \beta$ for 
$\,  {\tilde \Pi}$ and $\, {\tilde \Pi}_p$, because changing 
$\,  {\tilde \Pi}$ into $\, {\tilde \Pi}_p$, 
amounts to changing  $\, \nu$ into $\, 1/\nu$,
$\, S_0$ being invariant. Recalling  identity (\ref{thirdident}), a  
straightforward calculation shows that this identity is preserved 
by a representation of the duality 
$\, (k, \, K, \, {\tilde \Pi}, \, {\tilde \Pi}_p) \, \rightarrow \, 
(1/k, \, k \cdot \, K, \, {\tilde \Pi}^{(d)}, \, {\tilde \Pi}_p^{(d)})$, 
{\em only} for $\, \alpha \, = \, \, 1$ and $\, \beta\, = \, 0$. 

Thus, the representation of the Kramers-Wannier 
duality (\ref{Kramers}), very simply extends to the two  
complete elliptic integrals of the third kind $\, {\tilde \Pi}$ 
and $\, {\tilde \Pi}_p$ as follows:
\begin{eqnarray}
\label{Kramers2}
\hspace{-0.95in}&&  \, \, \quad      \quad  
(k, \, \, {\tilde E}, \,\, {\tilde K}, 
\, \, {\tilde \Pi}, \,\, {\tilde \Pi}_p) 
\, \quad \longrightarrow \quad \quad 
\Bigl( {{1} \over {k}}, \, \, {\tilde E}_{d}, \,\,  {\tilde K}_{d}, 
\,\, {\tilde \Pi}^{(d)}, \, \,{\tilde \Pi}_p^{(d)}\Bigr)
\nonumber \\
\hspace{-0.95in}&&  \, \, \quad   \quad  
\quad   \quad \quad  \quad     \quad 
 \, \, \,  \,  = \, \,  \,    \,   \, 
\Bigl( {{1} \over {k}}, \, \, \,
{{{\tilde E}} \over {k}} \, 
+ \, \, {{k^2 \, -1} \over {k}} \cdot \, {\tilde K},
 \,\,\,  k \cdot \, {\tilde K}, \,\, \, 
 k \cdot \, {\tilde \Pi}, \,\, \,  k \cdot \, {\tilde \Pi}_p \Bigr). 
\end{eqnarray}
or, in terms of the variables $\, s_h$,  $\, s_v$:
\begin{eqnarray}
\label{Kramers3}
\hspace{-0.95in}&&  \, \,    
(s_h, \,\, s_v, \,\, {\tilde E}, \,\, {\tilde K}, 
\, \,{\tilde \Pi}, \,\, {\tilde \Pi}_p) \,\,\,
 \quad \longrightarrow \quad \quad 
\Bigl( {{1} \over {s_v}}, \,\, {{1} \over {s_h}}, 
\,\,  E_{d}, \, \, K_{d}, \,\, 
{\tilde \Pi}^{(d)}, \,\, {\tilde \Pi}_p^{(d)}\Bigr)
\nonumber \\
\hspace{-0.95in}&&  \, \,  
 \, = \, \,   \, 
\Bigl( {{1} \over {s_v}}, \,\, {{1} \over {s_h}},  \,  \,
{{ {\tilde E}} \over {s_h \, s_v}} \, 
+ \, \, {{s_h^2 \, s_v^2 \, -1} \over {s_h \, s_v}} \cdot \, {\tilde K}, 
\,\, \,  s_h \, s_v \cdot \, {\tilde K}, \, \, \,
 s_h \, s_v \cdot \, {\tilde \Pi}, \, \, \, 
s_h \, s_v \cdot \, {\tilde \Pi}_p \Bigr). 
\end{eqnarray}
It is straightforward to verify that 
{\em this transformation is, as it should, an involution}. 

\vskip .2cm

\section{Linear differential operators for the anisotropic correlation functions}
\label{diffoperapp}

For simplicity we restrict here to  row correlation functions
$\, C(0, \, N)$, but everything  can be easily generalized 
to general $\, C(M, \, N)$ correlation functions.

Let us introduce
\begin{eqnarray}
\label{Introduce2}
\hspace{-0.95in}&&  \, \, \, \,  \,  \,  \quad \quad \quad \quad 
w \, \, = \, \, \, 
 \Bigl(((1 \, +s_h^2)\, (1 \, + s_v^2\Bigr)^{1/2}  
\, \, = \, \, \,  \, 
    \Bigl((1 \, +\nu \, k)\, (1\, +{{k} \over {\nu}})\Bigr)^{1/2}.
\end{eqnarray}
and the order-one partial differential operator  $\, M_1$:
\begin{eqnarray}
\label{Introduce3}
\hspace{-0.95in}&&  \, \, \quad \quad \quad \quad 
M_1 \, \, \, = \,  \,\, \, 
  {{1} \over {w}} \cdot \, 
  \frac{\partial }{\partial k}   \cdot \, w 
\, \,\,  \, = \, \, \, \,   \,\, 
 \frac{\partial }{\partial k}   \, \, \, 
+ \, {{\nu^2 \, +2\,\nu\, k \, +1} \over { 
2 \cdot \, (1\, + \, \nu \, k) \cdot \, (k \, + \, \nu)}}.
\end{eqnarray}

The action of the order-one operator  $\, M_1$ on the complete 
elliptic integral of the third kind $\, {\tilde \Pi}$ 
gives a linear combination of $\, {\tilde E}$ and 
$\, {\tilde K}$ {\em without} $\, {\tilde \Pi}$: 
\begin{eqnarray}
\label{gives}
\hspace{-0.95in}&&  \, \, \, \,  \,  
M_1({\tilde \Pi}) \, \, = \, \,  \,\, 
-\, {{{\tilde K}} \over { 2 \cdot \, (1\, +\nu \, k) \cdot \, k}} 
\, \, \, \,
 -\, {{ \nu \cdot \, (k^2+1) \, + \, 2 \, k} \over { 
2 \cdot \, (1\, +\nu \, k) \cdot \, (k \, +\nu) 
\cdot \, (k^2 \, -1)}} \cdot {\tilde E}.
\end{eqnarray}

More generally, let us consider a linear differential 
operator of order $\, N$: 
\begin{eqnarray}
\label{IntroduceL}
\hspace{-0.95in}&&   \quad \quad 
L_N  \,  \, \, = \, \, \,  \,  
p_N(k, \,\nu) \cdot \,    \frac{\partial^N }{\partial k^N} 
 \,  \,\,  
 + p_{N-1}(k, \,\nu) \cdot \,  \frac{\partial^{N-1} }{\partial k^{N-1}} 
\, \,  \,  
+  p_{N-2}(k, \,\nu)\cdot \,  \frac{\partial^{N-2} }{\partial k^{N-2}} 
\nonumber \\
\hspace{-0.95in}&&   \quad  \quad \quad \quad \quad \quad \quad 
 \, + \,  \, \, \, \cdots 
\, \,  \,  
+ \, \, p_{1}(k, \,\nu) \cdot \,  \frac{\partial}{\partial k} 
\, \,\, + \, \, p_{0}(k, \,\nu),
\end{eqnarray}
the action of its conjugate, 
$\, 1/w^m \cdot \, L_N  \cdot \, w^m$, 
on the product of a function $\, f(k)$ with
 $\, {\tilde \Pi}^m$, the $\, m$-th power of the complete 
elliptic integral of the third kind, gives:
\begin{eqnarray}
\label{IntroduceLconj}
\hspace{-0.95in}&&   \quad  
\Bigl({{1} \over {w^m}} \cdot \, L_N  
\cdot \, w^m\Bigr) \Bigl(f(k) \cdot \, {\tilde \Pi}^m\Bigr)  
 \, \, = \, \, \,  
\nonumber \\
\hspace{-0.95in}&&   \quad     \quad 
\, \, = \,\, \, \,  
L_N(f_0(k)) \cdot \, {\tilde \Pi}^{m} \,\, 
+ \, \,  \, {\cal P}_1(E, \, K) \cdot {\tilde \Pi}^{m-1}
 \, \,+ \, \,  \, 
{\cal P}_2(E, \, K) \cdot {\tilde \Pi}^{m-2}
 \,\,  + \, \,  \, \cdots 
\end{eqnarray}
and, of course, if $\, f_0(k)$ is solution of the linear 
differential operator (\ref{IntroduceL}) one gets:
\begin{eqnarray}
\label{IntroduceLconjmore}
\hspace{-0.99in}&&    
\Bigl({{1} \over {w^m}} \cdot \, L_N  \cdot \, w^m\Bigr)
 \Bigl(f_0(k) \cdot \, {\tilde \Pi}^m\Bigr)  
   = \, \,    {\cal P}_1(E, \, K) \cdot {\tilde \Pi}^{m-1}
  \,+   \, {\cal P}_2(E, \, K) \cdot {\tilde \Pi}^{m-2} 
\, +  \, \cdots 
\end{eqnarray}
and, more generally:
\begin{eqnarray}
\label{IntroduceLconj2}
\hspace{-0.95in}&&    \quad  \quad \quad \quad 
\Bigl({{1} \over {w^m}} \cdot \, L_N  \cdot \, w^m\Bigr) 
\Bigl(f_0(k) \cdot \, {\tilde \Pi}^m \, 
+ \, \, f_1(k) \cdot \, {\tilde \Pi}^{m-1} \, + \, \, \cdots  \Bigr)  
\nonumber \\
\hspace{-0.95in}&&  \quad \quad \quad \quad 
\quad  \quad  \quad   \quad  \quad       
 \,  = \, \,  \,  \,  \,   
 {\cal P}_1({\tilde E}, \, {\tilde K}) \cdot {\tilde \Pi}^{m-1}
  \, \, +   \, {\cal P}_2({\tilde E}, \, {\tilde K})
 \cdot {\tilde \Pi}^{m-2} \,  \,  \, + \,  \, \cdots 
\end{eqnarray}

Let us now apply equation (\ref{IntroduceLconj2}) on
 the row correlation functions $\, C(0, \, N)$ are 
homogeneous polynomials in $\,  {\tilde E}$, $\,  {\tilde K}$ 
and $\, {\tilde \Pi}$,
with coefficients that are rational functions in $\, k$ 
and $\, \nu$ (up to overall square root,  
$\, (s_v^2 \, +1)^{1/2} \, = \,\,  (1+ \, k/\nu)^{1/2} \, $ 
for $\, N$ odd). The row correlation functions 
$\, C(0, \, N)$ are of the form 
\begin{eqnarray}
\label{IntroduceLconj2row}
\hspace{-0.95in}&&    
\quad \rho_0(k, \, \nu) \cdot \, {\tilde \Pi}^m \, 
+ \, \, \rho_1(k, \, \nu, \, E, \, K) \cdot \, {\tilde \Pi}^{m-1} \, 
+  \, \rho_2(k, \, \nu, \, E, \, K) \cdot \, {\tilde \Pi}^{m-2} 
 \, + \, \,   \cdots 
\end{eqnarray}
where the $\, C_n$ are homogeneous polynomials in 
$\, {\tilde E}$ and $\, {\tilde K}$  
of degree $\, \nu$ in $\, {\tilde E}$ and 
$\, {\tilde K}$. The first coefficient $\,C_0(k, \, \nu)$ 
is annihilated by a first order linear differential operator 
\begin{eqnarray}
\label{firstM1}
\hspace{-0.95in}&&  \quad \quad  \quad \quad   \quad 
M_1 \, \, \, = \, \,  \, \,  {{\partial }  \over {\partial k}}
 \, \, \, - \, { {\partial \ln(C_0(k, \, \nu)} \over {\partial k}}, 
\end{eqnarray}
and one thus has, from  (\ref{IntroduceLconj2}), that
\begin{eqnarray}
\label{M1onC0N}
\hspace{-0.95in}&&  \quad \quad \quad \quad 
X_1 \, \, = \, \, \, 
\Bigl({{1} \over {w^m}} \cdot \, M_1  \cdot \, w^m\Bigr) (C(0, \, N)) 
\nonumber \\
\hspace{-0.95in}&&  \quad \quad \quad \quad \quad  \quad \quad 
 \,  = \, \,  \,  \,  \,  \,   
 {\cal P}_1({\tilde E}, \, {\tilde K}) \cdot {\tilde \Pi}^{m-1}
  \, \,  \, +   \, {\cal P}_2({\tilde E}, \, {\tilde K})
 \cdot {\tilde \Pi}^{m-2} 
\,  \, \,  \, + \,  \,  \, \cdots 
\end{eqnarray}
where the $ {\cal P}_n(E, \, K)$ are 
{\em also homogeneous polynomials in} $\, {\tilde E}$ 
{\em and} $\, {\tilde K}$ of degree $\, n$ in 
$\, {\tilde E}$ and $\, {\tilde K}$.
Let us denote $\, L_K$ the second order operator that 
annihilates $\, K$. At the next step, since 
$\, {\cal P}_1(E, \, K)$ is a linear combination of 
$\, {\tilde E}$ and $\, {\tilde K}$, one knows that 
there exists a second order operator $\, M_2$, homomorphic 
to $\, L_K$, that annihilates 
$\, {\cal P}_1({\tilde E}, \, {\tilde K})$. 
Consequently, using (\ref{IntroduceLconj2}), one gets that
\begin{eqnarray}
\hspace{-0.95in}&&  \quad \quad \quad \quad  
X_2 \, = \, \, 
\Bigl({{1} \over {w^{m-1}}} \cdot \, M_2  \cdot \, w^{m-1}\Bigr) (X_1) 
\\
\hspace{-0.95in}&&  \quad \quad \quad \quad \quad  \quad \quad 
 \,  = \, \,  \,  \,  \,  \,     \, 
{\cal Q}_2({\tilde E}, \, {\tilde K}) \cdot {\tilde \Pi}^{m-2} 
 \,\, + \, {\cal Q}_2({\tilde E}, \, {\tilde K}) \cdot {\tilde \Pi}^{m-3} 
\,  \,  \, + \,  \,  \, \cdots 
\end{eqnarray}
Again $\, {\cal Q}_2({\tilde E}, \, {\tilde K})$ being a homogeneous 
quadratic polynomial in $\, {\tilde E}$ and $\, {\tilde K}$, there 
exists an order-three linear differential operator $\, M_3$, 
homomorphic to the symmetric square of $\, L_K$,
that annihilates $\, {\cal Q}_2({\tilde E}, \, {\tilde K})$, and 
so on ... the last operator $\, M_{N+1}$, annihilating a homogeneous 
polynomial in $\, E$ and $\, K$ of degree $\, N$  
in $\, {\tilde E}$ and $\, {\tilde K}$,
 being homomorphic to the symmetric $\, N$-th power of $\, L_K$,
and of order $\, N+1$.

One immediately deduces a canonical decomposition
of the linear differential operator (in $\, k$, $\, n$ is fixed)
annihilating $\, C(0, \, N)$ in the form  
\begin{eqnarray}
\label{canonical}
\hspace{-0.95in}&&  \quad \quad \quad  \quad \quad \quad \quad 
 M_{N+1} \cdot \,  M_{N} \cdot \, M_{N-1} 
\, \,  \, \cdots \,\,   \, \, M_{2} \cdot \, M_{1},  
\end{eqnarray}
where the $\, M_n$'s are homomorphic to the $\, (n-1)$-th symmetric power 
of $\, L_K$. One recovers that the (partial) 
linear differential operator in $\, k$
($\, \nu$ is fixed) annihilating $\, C(0, \, N)$ 
is of order $\, (N+1) \, (N+2)/2$.

\vskip .2cm 

\vskip .1cm 

{\bf Remark 1:} Similar calculations can be performed on the  $\, C(M, \, N)$, 
mutatis mutandis. For instance, if one considers the exact expression for 
$\, C(1, \, 2)$, namely (\ref{offcorrel}), one sees that 
there is no $\, {\tilde \Pi}^2$ terms, but it is of the form 
$\, \rho_1(k, \, \nu, \, {\tilde E}, \, {\tilde K})
 \cdot \, {\tilde \Pi} \, +  \, \rho_2(k, \, \nu, \, {\tilde E}, \, {\tilde K})$, 
therefore one will have a decomposition of the form $\, M_3 \cdot \, M_2$
and not $\, M_3 \cdot \, M_2\cdot \, M_1$. 

\vskip .2cm 

\vskip .1cm 

{\bf Remark 2:}  The previous calculations where performed in the 
$\, k$, $\, \nu$ variables, but there is {\em nothing specific with 
these variables}. One obtains similar results in $\, s_h$ 
and $\, s_v$. For instance, seen as  functions of $\, s_h$, $\, s_v$ 
being fixed, the row correlation $\, C(0,\, 2)$ is, again, solution 
of a linear differential operator of order six,
which factorizes as $\, M_3 \cdot \, M_2 \cdot \, M_1$, 
 $\, C(1,\, 3)$ is solution of a linear differential operator 
of order nine, which factorizes as $\, M_4 \cdot \, M_3 \cdot \, M_2$, 
and, more generally $\, C(M, \, N)$ is solution of a linear 
differential operator of order $\, (N\, -N\, +1) (N\, +M\, +2)/2$, 
which factorizes as $\,M_{N+1} \cdot \,  M_{N} \cdot \, M_{N-1} 
\, \,  \, \cdots \,\,   \, \, M_{M\, +2} \cdot \, M_{M\,+1}$.

\vskip .2cm 

\subsection{Linear differential operators for the anisotropic correlation functions}
\label{diffopersub}

 The previous calculations can of course be performed in a similar way for 
the low-temperature correlations, mutatis mutandis, yielding the same  
order $\, (N+1) \, (N+2)/2$ for the operator for $\,C_<(0,N)$, with the same
canonical factorization (\ref{canonical}) on the corresponding operators.

For example the low-temperature correlation function  $\,C_<(0,1)$ is annihilated 
by an order-three linear differential operator $\, M_3\,\, =\,\,\,M_2\cdot M_1$, 
where setting $ \, x\,=\,\,k_<$ the order-one and order-two 
operators $\, M_1$ and $\, M_2$ read: 
\begin{eqnarray}
\hspace{-0.97in}&& \quad \quad \, \, 
M_1 \,\,=\,\, \,  {{\partial } \over { \partial x}} \, \, -\frac{1}{2 \,(x+\nu)}, 
\quad \quad \, \, 
M_2 \,   \, =\, \,  \,  \,
 p_2\cdot  \,  {{\partial^2 } \over { \partial x}}  \,\,
 +p_1\cdot  \, {{\partial } \over { \partial x}}  \,\, +p_0, 
\end{eqnarray}
with
\begin{eqnarray}
\hspace{-0.97in}&& \quad \quad \quad \quad
p_2\, \,  = \, \, \, 
4 \cdot \,x\cdot \, (x^2-1)\cdot \,(x+\nu)^2
\cdot \,(\nu x+1)^2(\nu x^3 +3x^2 +3\nu x +1), 
\nonumber\\
\hspace{-0.97in}&& \quad \quad \quad \quad
p_1\, \, = \,\,  \,  4 \cdot \, (x+\nu) \cdot \, (1\, +\nu x) 
\cdot \, \{3 \nu^2 x^7 +(2\nu^2 +14)\, \nu \cdot \, x^6
\nonumber\\
\hspace{-0.97in}&& \quad \quad \quad \quad \quad
+(24\nu^2 +9) \cdot \, x^5 \, \, 
+(12\nu^2 +18)\, \nu \cdot \, x^4 \, \, 
+(3\nu^2 +2) \cdot \,x^3
\nonumber\\
\hspace{-0.97in}&& \quad \quad \quad \quad \quad
-6 \cdot \, (\nu^2 +1) \,  \nu \cdot \,x^2 
\, -(6\nu^2 +3) \cdot \, x -2 \, \nu\}
\nonumber\\
\hspace{-0.97in}&& \quad \quad \quad \quad
p_0 \, \, = \, \, \, 3 \, \nu^3 \cdot \, x^8
  +(8\nu^2 +19) \, \nu^2 \cdot \, x^7 \,
 +(\nu^4 +80\nu^2 +18) \, \nu \cdot \, x^6
\nonumber\\
\hspace{-0.97in}&& \quad \quad \quad \quad \quad
 +(55\nu^2 +140) \, \nu^2 \cdot \,x^5 \, 
+(6\nu^4 +123\nu^2 +96) \, \nu \cdot \,x^4
\nonumber\\
\hspace{-0.97in}&& \quad \quad \quad \quad \quad
+(18\nu^4 +99\nu^2 +36) \cdot \,x^3 \,
 +(9\nu^4 +10\nu^2 +38) \, \nu \cdot \,x^2
\nonumber\\
\hspace{-0.97in}&& \quad \quad \quad \quad \quad
+(15\nu^4 -2\nu^2 -4) \cdot \, x \,\, + 8 \, (\nu^2 -1) \, \nu.
\end{eqnarray}

\end{document}